\documentclass[10pt]{article}
\usepackage[utf8]{inputenc}
\usepackage{blindtext}
\usepackage{multicol}
\usepackage{hyperref}
\usepackage{amsmath}
\usepackage{multicol}
\usepackage{graphicx}
\usepackage[dvipsnames]{xcolor}
\usepackage{float}
\usepackage{longtable}
\usepackage{lipsum}

\usepackage[style=nature]{biblatex}
\usepackage{amsfonts}
\usepackage[margin=.4in]{geometry}
\usepackage{caption}
\usepackage{subcaption}
\hypersetup{colorlinks,citecolor=magenta,linkcolor=blue,urlcolor=magenta}
\title{\bf{ Exploring the dynamics of coincident $f(Q)$ gravity in the presence of DBI-essence scalar field}}
\author{\bf Ratul Mandal$^1$\footnote{e-mail:{ratulmandal2022@gmail.com}(Corresponding Author)}~,~ Ujjal Debnath$^1$\footnote{e-mail: {ujjaldebnath@gmail.com}} ~and~
Anirudh Pradhan$^2$\footnote{e-mail :
{pradhan.anirudh@gmail.com}}\\
$^1$Department of Mathematics, Indian Institute of Engineering
Science\\ and Technology, Shibpur, Howrah-711103, India.\\
$^2$Centre for Cosmology, Astrophysics and Space Science,\\ GLA
University, Mathura-281406, U.P., India. }

\date{\today}
\begin{document}
\maketitle

\begin{abstract}
\noindent
In theoretical cosmology, symmetric teleparallel gravity or $f(Q)$ gravity based on nonmetricity tensor $Q$ has become an interesting alternative to General relativity in recent years. The present research paper contains a rigorous dynamical system analysis of coincident $f(Q)$ gravity in the presence of a generalized DBI essence scalar field. We have considered two different models of coincident $f(Q)$ gravity, such as power law model $f(Q)=Q+nQ^m$ and exponential model $f(Q)=Qe^{\frac{\beta Q_0}{Q}}$ respectively, where $n,m,\beta$ are constant parameter and $Q$ is the nonmetricity component. In this study, the generalized DBI essence scalar field acts as an additional dark energy component. After obtaining the field equation for the corresponding cosmological model, we employed several dynamical variables to form the dynamical system. The critical points of these dynamical systems are influenced by cosmological parameters and associated with particular epochs in the cosmological timeline. For different combinations of cosmological parameters, the critical points exhibit different cosmological eras, starting from the accelerated stiff matter era to late-time acceleration phenomena. The stability criteria of each critical point are studied by using linear stability theory, and the physical constraints on the cosmological parameters are also considered during this analysis. Furthermore, the current values of energy densities, deceleration parameters, and equation of state parameters obtained from the evolution diagram are compatible with observational data.     
\end{abstract}

\begin{multicols}{2}
  \section{Introduction}
  \noindent
  	In the last century, "General relativity" was the most celebrated theory of physics. Instead of considering gravity as an attractive force, Einstein considered gravity to be a geometrical effect of space-time, and this interpretation has prepared the fundamental field of modern cosmology. In 1996, the observational data from type Ia Supernovae (SNeIa) \cite{Riess_1998} \cite{SupernovaCosmologyProject:1998vns} disclosed that our universe is going through an accelerating phase. Later, Cosmic Microwave Background (CMB)\cite{Hinshaw_2013}\cite{Planck:2013kqc} and Baryonic Acoustic Oscillation (BAO)\cite{Spergel_2003}\cite{Anderson:2012sa} confirmed it .After the Big Bang, our spatially flat, homogenous universe passed through two accelerating expansion phases: cosmic inflation, which happened just after the Big Bang, and late time acceleration, which is what our universe is currently experiencing. These observational results sharply challenge our fundamental understanding of gravity, and researchers are compelled to search for a suitable theory that can explain the acceleration phenomenon.\\
   In the last two decades, two physical theories have become relevant in this framework: respectively, the Dark Energy and modified gravity theories. Dark energy, as its name suggests, is a mysterious, invisible matter component with positive energy density and engenders sufficient negative pressure to hold the gravitational effect and initiate late-time acceleration. The proper origin of dark energy is still questionable. The most elementary way of introducing dark energy is by adding a cosmological constant term $\Lambda$ in the right-hand side of the Einstein field equation, which acted as a source of dark energy and dark matter, and it satisfies the equation of state $\omega_\Lambda=\frac{p_\Lambda}{\rho_\Lambda}=-1$. This dark energy model is known as the $\Lambda\textrm{CDM}$ model (CDM stands for cold dark matter). Numerous observational datasets have proven the incredible efficacy of the $\Lambda\textrm{CDM}$ model. However, it suffers from some serious drawbacks, like cosmic coincidence problem \cite{PhysRevLett.82.896}\cite{PhysRevLett.85.4434}, which refers to coinciding the energy densities of dark energy and matter at the present epoch despite evolving in a different manner and the cosmological constant problem \cite{PADMANABHAN2003235}\cite{RevModPhys.61.1} pertains to the discrepancy between the observed and theoretically predicted value of vacuum energy density, suggesting an unsolved issue in reconciling quantum field theory predictions with observational data. These shortcomings of the $\Lambda\textrm{CDM}$ model trigger researchers to introduce the dynamical dark energy model, such as the scalar field model (Quintessence model \cite{WETTERICH1988668}\cite{Tsujikawa_2013}, K-Essence model\cite{PhysRevD.63.103510}\cite{PhysRevD.62.023511}, DBI Essence model \cite{PhysRevD.77.123508} ), dark fluid model ( Tachyon model \cite{PhysRevD.74.043528}, Chaplygin gas \cite{Debnath_2004}). Self-interacting potentials in the scalar field model are able to provide an ideal cosmic evolution that replicates the effect of a cosmological constant $\Lambda$ in the present era. The value of the equation of state parameter $\omega$ for the quintessence model lies in the range $-1\leq \omega \leq1$ and for accelerated expansion, we get $\omega\leq-\frac{1}{3}$.In the case of the phantom model, $\omega <-1$, which seems to be more accurate to the observational data provided by Plank collaboration \cite{PhysRevD.88.063501}\cite{Novosyadlyj_2014}.On the other hand, the Dirac Born Infield (DBI) model is an intriguing dark energy model based on string theory and can explain the accelerated expansion. The kinetic part of DBI action is noncanonical. In the manuscript \cite{PhysRevD.77.123508}, the authors mentioned that the kinetic part originates from the fact that the system's action is proportional to the volume traced out by the Brane during its motion. This volume is given by the square root of the induced metric, which automatically leads to a DBI kinetic term. The potential part emerges from the internal tension of the D-brane, which governs the geometrical characteristics of the warped throat region. Copeland et al.in  \cite{PhysRevD.81.123501} have studied the detailed dynamics of the DBI field for both the power law form and exponential form of warp factor and potential. In \cite{PhysRevD.80.123016}\cite{AHN2010181}, Ahn et al. investigated that the DBI scalar field can exhibit late time attractor behavior for minimum potential. The DBI-essence scalar field introduces additional dynamics that can significantly affect the behavior of the cosmological models. It can lead to rich and complex behaviors, such as late-time acceleration (mimicking dark energy) and alleviating fine-tuning problems. The coupling between the DBI-essence scalar field and the symmetric teleparallel $f(Q)$ gravity can lead to new and interesting cosmological phenomena that are not present in standard $\Lambda$CDM models or other modified gravity theories. \\
   As we have mentioned earlier, there is an alternative approach to the dark energy model to explain the accelerated expansion, which aims to modify the Einstein-Hilbert action of general relativity, known as modified gravity theory. In the modification of gravity theory, no exotic matter components are required to explain the acceleration phenomena. One can classify the modified gravity theories into three classes.  In the first one, the gravitational interaction is governed by the curvature of space-time. The simplest way of modifying Einstein gravity with the help of the curvature term is to substitute the Ricci Scalar term $R$ by a general function $f(R)$ in the Einstein-Hilbert action. This modification is known as $f(R)$ gravity theory\cite{DeFelice:2010aj}. Sotiriou et al.in \cite{RevModPhys.82.451} studied the cosmic evolution with the help of nonlinear curvature term. The dynamical aspect of modified $f(R)$ gravity is studied in the literature \cite{Shah:2020puj}\cite{PhysRevD.96.104049} \cite{Alho:2016gzi} \cite{Carloni:2015jla}, where the authors have considered several forms of  $f(R)$ and use a dynamical system approach to study the cosmological scenario. The black hole and wormhole solution under $f(R)$ gravity theory have been studied in the literature \cite{PhysRevD.80.124011}\cite{SHAMIR2021634} in recent years. Some other important modified gravity theories that fall into this class are modified $f(G)$ gravity theory \cite{NOJIRI20051} \cite{Carloni:2017ucm} , where $G$ is the Gauss-Bonnet invariant,$f(P)$ gravity theory \cite{PhysRevD.94.104005} \cite{PhysRevD.99.123527}where $P$ is a nontopological cubic invariant arising from the contraction of the Riemann tensor in a specific order. In the second class of modified gravity theories, instead of curvature, the gravitational interaction is guided by the torsion term $T$, known as the teleparallel equivalent of general relativity(\textbf{TEGR}).A generalization of teleparallel gravity has been formulated by substituting the torsion term $T$ with a general function $f(T)$\cite{PhysRevD.75.084031}\cite{doi:10.1142/S0217751X09045236}. An extensive investigation on several properties of modified $f(T)$ theory is presented in \cite{Myrzakulov:2010vz} \cite{PhysRevD.84.043527} \cite{Cai_2016} \cite{Mishra:2019vnv}.In recent times, another important class of modified gravity theory has been introduced, where the gravitational interaction is governed by a curvature-less and torsion-free nonmetric compatible connection $Q$, known as the symmetric teleparallel equivalent of general relativity (\textbf{STEGR}). These three different approaches for modifying gravity theory by using three different geometrical connections, such as curvature, torsion, and nonmetricity, respectively, are known as the "geometrical trinity of gravity."\\
   A generalization of symmetric teleparallel gravity, known as $f(Q)$ gravity \cite{PhysRevD.98.044048}\cite{PhysRevD.101.103507}, successfully satisfied various observational data like redshift space distortion, cosmic microwave background, baryonic acoustic oscillation, etc., which makes it an interesting alternative to the $\Lambda CDM$ model.In  $f(Q)$ gravity, we get a second-order field equation, while the field equation corresponding to usual curvature-based $f(R)$ gravity is of fourth order. According to Weyl's geometry,$f(Q)$ gravity is the conception of Riemannian geometry. The dynamical features of $f(Q)$ and $f(T)$ gravity are similar, but a significant difference originates in the perturbation level.An advantage of $f(Q)$ gravity over $f(T)$ gravity is that the cosmological solution provided by $f(Q)$ gravity can contain the exact solution obtained from general relativity, and it can also go beyond that. But for the case of $f(T)$ gravity, the solutions are completely fixed, and not dynamical \cite{DAmbrosio:2021pnd}. In $f(Q)$ gravity, the connection can be simplified by partial derivatives, which may vanish for some particular choices of coordinate known as the coincident gauge. In the case of non-coincident gauge, the connections are not constant and function of time. In this present article, we have completely focused on the coincident form of $f(Q)$ gravity. In the past few years, several aspects of $f(Q)$ gravity have been studied. The black hole and wormhole solution under $f(Q)$ gravity was studied in \cite{Bahamonde_2022}\cite{Javed:2023qve}\cite{sym13071260}\cite{Banerjee:2021mqk}.An analysis related to the cosmological constraint on noncoincident $f(Q)$ gravity is presented in \cite{Pradhan:2024eew} .In \cite{PhysRevD.102.024057}, Mandal et al. have investigated the energy conditions for the power law model and logarithmic form of $f(Q)$ gravity to determine the stability of these cosmological models, and the authors have also constrained the cosmological parameters for current observational values. Moreover, a brief analysis of the singularities of $f(Q)$ gravity has been studied in \cite{PhysRevD.103.044021}.In \cite{PhysRevD.98.084043}, Harko et al. considered an extension of $f(Q)$ gravity by incorporating nonminimally coupled matter lagrangian with nonmetricity $Q$ , resulting to  nonconservation of the energy-momentum tensor.Some other important aspects of modified $f\left(Q\right)$ gravity has been studied intensively in the following literature \cite{Pradhan:2021qnz},\cite{Pradhan:2022dml},\cite{Maurya:2023fgy},\cite{Goswami:2023knh}.Although the $f(Q)$ gravity receives significant attention in cosmology and black hole physics, it suffers from some serious problems revealed by some recent studies. These class of gravity theories are found to be infinitely strongly coupled in the neighborhood of maximally symmetric background \cite{PhysRevD.101.103507},\cite{universe7050143}. Additionally, the appearance of expected ghosts is a serious hazard for these classes of theories.In \ cite {PhysRevLett.132.141401}, the authors examined the reason behind the appearance of ghosts in symmetric teleparallel gravity. The authors also revealed the existence of seven dynamical degrees of freedom corresponding to the gravitational sector in one of the branches of $f(Q)$ gravity, and at least one of them is a ghost.\\
   The implementation of the dynamical system approach in cosmology has become famous in the last few years \cite {BAHAMONDE20181}\cite{doi:10.1142/S021827180600942X}. The dynamical system analysis method provides useful mathematical techniques to handle the nonlinearity of field equations\cite{coley1999dynamicalsystemscosmology}\cite{Boehmer:2014vea}. The critical points of the dynamical systems indicate a particular epoch in the cosmological timeline. By assigning the proper variables and tracing the evolution of the corresponding variable, several dynamical aspects of a particular cosmological model can be understood briefly. Moreover, the stability properties corresponding to each critical point can be analyzed in detail by applying linear stability theory and center manifold theory according to the nature of the critical points. Dynamical system analysis of DBI essence scalar field presented in \cite{Pal:2019qch}\cite{doi:10.1142/S0217732315500091} \cite{Bhadra:2012qj}. Dynamical system analysis $f(Q)$ gravity is presented in the following literature \cite{PhysRevD.107.044022}\cite{PhysRevD.103.103521}\cite{NARAWADE2022101020}\cite{Vishwakarma_2023}\cite{universe9040166}\cite{Vishwakarma:2024qvw}\cite{Ghosh_2024}. In this present article, we have presented a brief analysis on the dynamics of $f(Q)$ gravity in the presence of generalized DBI scalar field. We have investigated two different forms of $f(Q)$ gravity, such as the power law model $f(Q)=Q+m Q^n$ and the exponential model $f(Q)=Qe^{\frac{\beta Q_0}{Q}}$ respectively, where $\left(m, n, \beta \right)$ are constant parameter. The dynamical system corresponding to each cosmological model is formulated by introducing some dimensionless variables. The dynamical stability behavior of each critical point has been studied in detail by applying linear stability theory. We have also tried to find out the constraint on the parameter to exhibit stable behavior and satisfy the current cosmological scenario. We have organized this article in the following manner: In section \ref{Sec2}, the basics of $f(Q)$ gravity and generalized DBI essence are presented; in sec \ref{Sec3}, we have introduced dynamical variables and constructed the dynamical system. The next two subsections contain the table of critical points, the detailed analysis for each critical point, and the simulation of phase trajectories. Finally, the paper ends with a valuable conclusion in section \ref{sec4}.
   
\section{Background of symmetric teleparallel $f(Q)$ gravity}\label{Sec2}
The metric tensor $g_{\mu \nu}$ in Riemannian geometry governs the geometry of space-time, and the affine connection $\Gamma^\lambda _{\mu \nu}$ indicates the parallel transport. In the General theory of relativity, the Levi-Civita connection is a symmetric connection that governs the gravitational interaction. Now, the general metric affine connection can be decomposed into three independent components as\cite{HEHL19951}\\
\begin{eqnarray}\label{eq1}
    \Gamma^\lambda _{\mu \nu}=\{^\lambda _{\mu \nu}\}+ K^\lambda_{\mu\nu} +L^\lambda_{\mu \nu}
\end{eqnarray}
The first component of\eqref{eq1} represents the Levi-Civita connection of $g_{\mu\nu}$ , and its defined by 
\begin{eqnarray}
    \{^\lambda _{\mu \nu}\}=\frac{1}{2}g^{\lambda \alpha} \left(\partial_{\mu} g_{\mu \alpha}+\partial_{\nu} g_{\alpha \mu} -\partial_{\alpha} g_{\mu \nu} \right)
\end{eqnarray}
Here, $g^{\lambda \alpha}$ represents the inverse of metric tensor $g_{\lambda\alpha}$ .\\
The second components of \eqref{eq1} $K^\lambda_{\mu\nu}$ represents the contortion and it is defined by 
\begin{eqnarray}
    K^\lambda_{\mu\nu}=\frac{1}{2}T^\lambda _{\mu\nu}+T_{(\mu \;\;\;\nu)}^{\;\;\;\lambda}
\end{eqnarray}
Here, $T^\lambda_{\mu\nu}$ is the torsion tensor, which is defined as the anti-symmetric component of affine connection \\
\begin{eqnarray}
 T^\lambda _{\mu\nu}= 2\Gamma^\lambda_{[\mu \nu]}    
\end{eqnarray}
The third component of the right-hand sight of \eqref{eq1} represents disformation, and it is defined by \\
\begin{eqnarray}\label{eq5}
    L^\lambda_{\mu\nu}=\frac{1}{2} g^{\lambda\sigma}\left(-Q_{\mu\sigma\nu}-Q_{\nu\sigma\mu} +Q_{\sigma\mu \nu}\right) 
\end{eqnarray}
Where the non metricity tensor $Q_{\sigma \mu \nu}$ is defined as $Q_{\sigma\mu\nu}=\nabla _{\sigma} g_{\mu\nu}$.Now the nonmetricity scalar $Q$ is defined as 
\begin{equation}
    Q=-\frac{1}{4}Q_{\alpha\beta\gamma}Q^{\alpha\beta\gamma}+\frac{1}{2}Q_{\alpha\beta\gamma}Q^{\gamma\beta\alpha}+\frac{1}{4}Q_\alpha Q^\alpha-\frac{1}{2}Q_\alpha \Tilde{Q^\alpha}
\end{equation}
Where, $Q_{\alpha}=  Q_{\alpha\;\mu}^{\;\;\mu}$ and $\Tilde{Q}^{\alpha}=Q_\mu ^{\;\;\mu\alpha}$ are two independent traces of nonmetricity tensor.\\
Using the definition of nonmetricity tensor $Q_{\sigma \mu \nu}$,from \eqref{eq5} the expression of disformation tensor can be written as\\
\begin{eqnarray}
    L^\lambda _{\mu\nu}=\frac{1}{2}\left(-\nabla_\mu g_{\sigma \nu} -\nabla_\nu g_{\sigma \mu} +\nabla_\sigma g_{\mu\nu}\right)
\end{eqnarray}
Now the curvature of space-time can be determined by  the Riemann tensor $R^\sigma _{\beta \mu\nu}$,we can define the Riemann tensor in terms of the contortion and disformation tensors as:
\begin{multline}\label{eq8}
    R^\lambda _{\beta\mu\nu}=\mathring{R}^\lambda _{\beta\mu\nu}+\mathring{\nabla}_\mu \left(K^\lambda _{\nu\beta}+L^\lambda_{\nu\beta}\right)-\mathring{\nabla}_\nu \left(K^\lambda _{\mu\beta}+L^\lambda_{\mu\beta}\right)+ \\
    \left(K^\lambda _{\mu\alpha}+L^\lambda_{\mu\alpha}\right)\left(K^\alpha _{\nu\beta}+L^\alpha_{\nu\beta}\right)-\left(K^\lambda _{\nu\alpha}+L^\lambda_{\nu\alpha}\right)\left(K^\alpha _{\mu\beta}+L^\alpha_{\mu\beta}\right)
\end{multline}
Here, $\mathring{R}^\lambda _{\beta\mu\nu}$ and $\mathring{\nabla}_\mu$ are presented in terms of Levi-Civita connection. Now by considering torsion free constraint $T^\lambda _{\mu\nu} = 0$ and appropriate contraction to the curvature term ,from \eqref{eq8} we get\\
\begin{eqnarray}\label{eq9}
    R=\mathring{R} - Q +\mathring{\nabla}_\lambda \left(Q^\lambda - \Tilde{Q}^\lambda \right)
\end{eqnarray}
Here $\mathring{R}$ is Ricci scalar calculated in terms of Levi-Civita connection, and to obtain curvature less teleparallel gravity, we apply the constraints $R=0$, then from \eqref{eq9} we get \\
\begin{eqnarray}\label{eq10}
    \mathring{R} = Q + \mathring{\nabla}_\lambda \left(Q^\lambda - \Tilde{Q}^\lambda \right)
\end{eqnarray}
The relation found in equation \eqref{eq10} shows that there is a boundary term separating the non-metricity scalar and the Ricci scalar curvature. The gravity theory only takes into account non-metricity scalar Q in the action, which is different from Einstein's GR by a boundary term, as shown by equation \eqref{eq10}. Because of this, STEGR is referred to as the symmetric teleparallel equivalent of GR theory since it offers a formulation that is comparable to GR.\\
In the present work, we have also considered a DBI-essence scalar field as an additional dark energy component. Now the action for $f(Q)$ gravity in the presence of DBI-essence scalar field is \\
\begin{multline}
    S=\int{\frac{1}{2} f(Q)\sqrt{- g} \hspace{0.2cm}d^4 x}+\int{L_{DBI} \sqrt{- g}\hspace{0.2cm} d^4 x}\\+ \int{L_M \sqrt{- g} \hspace{0.2cm}d^4 x}\hspace{3.89cm}
\end{multline}
Here, $f(Q)$ is arbitrary function of nonmetricity scalar $Q$ and $g=det\left(g_{\mu\nu}\right)$, $L_M$ is lagrangian for matter part and $L_{DBI}$ is Lagrangian corresponding to the DBI essence given by \cite{Pal:2019qch}\\
\begin{eqnarray}\label{eq12}
    L_{DBI}=-\left(\frac{1}{f\left(\phi\right)}\left(\sqrt{1-2 f\left(\phi\right) X}-1\right)-V\left(\phi \right)\right)
\end{eqnarray}
Where,$X=-\frac{1}{2} g^{\mu \nu}\partial_\mu \phi \partial_\nu \phi$\\
\\
In the expression of $L_{DBI}$ in \eqref{eq12}, $V\left(\phi\right)$ represents the potential of DBI field, which arises from the quantum interaction of D3 brane and the inverse of D3- brane tension is representing by $f\left(\phi\right)$ called the "Warp factor". In this article, we have considered that both the potential and warp factors are positive and are in exponential form i.e $V\left (\phi\right)=e^{\lambda\phi}$ and $f\left(\phi\right)=e^{\mu\phi}$, where $\lambda,\mu$ are constant parameters.

The expression for the stress-energy tensor of the\hspace{2cm} DBI scalar field is \\
\begin{eqnarray}\label{eq13}
    T^\phi _{\mu\nu}=-\frac{2}{\sqrt{-g}} \frac{\delta L_{DBI}}{\delta g^{\mu\nu}}
\end{eqnarray}
We know that for a perfect fluid, stress-energy tensors have the expression\\
\begin{eqnarray}\label{eq14}
    T_{\mu\nu}=\left(\rho+p\right)u_\mu u_\nu +p g_{\mu\nu}
\end{eqnarray}
Where $u^\nu=(1,0,0,0)$ are the usual four velocity vectors.N ow by comparing \eqref{eq13}and \eqref{eq14} we get the energy density $\rho_{\phi}$ and pressure $p_{\phi}$ for DBI scalar field as:\\
\begin{eqnarray}
    \rho_{\phi}=\frac{\nu^2}{\nu + 1}\Dot{\phi}^2+V\left(\phi\right)\label{eq15}\\
    p_{\phi}=\frac{\nu}{\nu +1}\Dot{\phi}^2-V\left(\phi\right)\label{eq16}
\end{eqnarray}
Here "." represents derivative with respect to time t, and $\nu$ is analogous to the Lorentz boost factor \cite{Pal:2019qch}, given by 
$\nu=\frac{1}{\sqrt{1-f\left(\phi\right) \Dot{\phi}^2}}$\\
\\
From the above expression, one can note that for the existence of $\nu$ in real physical space, the necessary constraint is  $f\left(\phi\right)\Dot{\phi}^2 \leq 1$ and therefore $\nu\geq 0$\\
The energy balance equation for DBI field is \\
\begin{eqnarray}
    \Dot{\rho_\phi} +3H\left(\rho_\phi +p_\phi \right)=0\label{eq17}
\end{eqnarray}
$H=\frac{\Dot{a}}{a}$ is the usual Hubble parameter and $a\left(t\right)$ is the scale factor. Substituting the value of $\rho_\phi$ and $p_\phi$ from \eqref{eq15} and \eqref{eq16} respectively in \eqref{eq17} we get the modified Klein-Gordon equation for DBI field as\cite{Pal:2019qch}:\\
\begin{multline}\label{eq18}
   \frac{2\nu^2}{\nu +1} \Dot{\phi} \Ddot{\phi}+\left(\frac{2\nu}{\nu+1}-\frac{\nu^2}{\left(\nu+1\right)^2 }\right)\Dot{\nu} \Dot{\phi}^2+\frac{dV}{d \phi}\Dot{\phi}\\+3H\nu\Dot{\phi}^2=0 \hspace{5.5cm} 
\end{multline}
Now our next step is to obtain the field equation for modified $f(Q)$ gravity in the presence of DBI field, and for this, we choose the background as flat, homogeneous, isotropic Robertson-Walker geometry given by the following metric\\
\begin{eqnarray}\label{eq19}
    ds^2=-dt^2+a^2\left(t\right) \left(dx^2+dy^2+dz^2\right)
\end{eqnarray}
Due to the line element \eqref{eq19} we can write down the trace of the non-metricity scalar $Q$ as $Q=6H^2$.We split the function $f\left(Q\right)$ as $f\left(Q\right)=Q+\Psi\left(Q\right)$.Where $\Psi\left(Q\right)$ is an arbitrary function of non-metricity scalar $Q$.If one take $\Psi\left(Q\right)=0$, we can immediately recover the Einstein GR from $f\left(Q\right)=Q$.Therefore for the previously mentioned metric \eqref{eq19}  the field equation for modified coincident $f\left( Q\right)$ gravity in the presence of DBI scalar field is\cite{PhysRevD.98.044048}\cite{PhysRevD.101.103507} \\
\begin{eqnarray}
    3H^2=\rho+\frac{\Psi}{2}-Q \Psi_{Q}\hspace{2cm}\label{eq20}\\
    \left(2 Q \Psi_{Q Q} + \Psi_Q+1\right) \Dot{H} + \frac{1}{4} \left( Q + 2 Q \Psi_Q - \Psi \right) = -2 p\label{eq21}
\end{eqnarray}
Here , $\rho=\rho_M + \rho_\phi$ and $p=p_M + p_\phi$\\.
Also, $\rho_M$ is the matter density and $p_M$ is the pressure component corresponding to the dark matter sector, respectively.  In this work, we have considered a dust form of dark matter, which will lead us to $p_M =0$.On the other hand, $\rho_\phi$ and $p_\phi$ are the energy density and pressure for the DBI scalar field, respectively. Hence we can finally write down the field equation \eqref{eq20} and \eqref{eq21} in a much simpler form:\\
		\begin{eqnarray}
		3H^2=\rho_M +\rho_\phi+\rho_Q\label{eq22}\\
		2\Dot{H}+3H^2=-p_\phi-p_Q\label{eq23}
		\end{eqnarray}
		 Here, $\rho_Q$ and $p_Q$ are energy density and pressure components that arise from geometrical dark energy having the following expression:\\
		 \begin{eqnarray}
		 	\rho_Q=\frac{\Psi}{2}-Q \Psi_{Q}\\
		 	p_Q= 2\Dot{H}\left( 2 Q \Psi_{Q Q} +\Psi_{Q}\right) -\rho_Q
		 \end{eqnarray}
   In this work, we have considered that the dark matter and dark energies do not interact with each other, and hence by applying the fact that stress-energy tensors are divergence-less leads us to the energy balance equations for dark matter and geometrical dark energies are given as\\
		 \begin{eqnarray} 
		 	\Dot{\rho_M}+3H\rho_M=0\hspace{1cm}\label{eq26}\\
		 	\Dot{\rho_Q}+3 H \left( \rho_Q +p_Q\right) =0\label{eq27}
		\end{eqnarray}
		To study the cosmological scenario we introduce the density parameters corresponding to dark matter, DBI field and geometrical dark energy component, respectively as\\
		\begin{eqnarray}\label{eq28}
			\Omega_M = \frac{\rho_M}{3H^2},\Omega_\phi=\frac{\rho_\phi}{3H^2},\Omega_Q =\frac{\rho_Q}{3H^2}
		\end{eqnarray}
		Using the density parameters in \eqref{eq28} the field equation \eqref{eq22} is transformed as\\
		\begin{eqnarray}\label{eq29}
		\Omega_M +\Omega_\phi+\Omega_Q=1
		\end{eqnarray} 
		From \eqref{eq29} we can identify a cosmological solution as completely dark energy dominated if $\Omega_Q=1$ and $\Omega_M=\Omega_\phi=0$.Similarly, a universe can be classified as matter-dominated or scalar field-dominated if $\Omega_M$ or $\Omega_\phi$, respectively, dominates over the others.\\
    
		The effective equation of state parameter is defined as\\
		\begin{eqnarray}
			\omega_{eff}=\frac{p_\phi +p_Q}{\rho_M+\rho_{\phi}+\rho_Q}=-1-\frac{2 \Dot{H}}{3 H^2}
		\end{eqnarray}
		The different phases of cosmic evolution are exhibited by the values of $\omega_{eff}$. For instance, $\omega_{eff}=-1$ denotes a de-sitter phase of evolution and dark energy components mimic the cosmological constant behavior on the other hand $\omega_{eff}=0$ signifies a matter-dominated universe and   $-1<\omega_{eff}<-\frac{1}{3}$ represents quintessence phase of evolution.\\
		The deceleration parameter can be defined in a similar way as\\
		\begin{eqnarray}
		q=-1-\frac{\Dot{H}}{H^2}=\frac{1}{2}\left( 1+3\omega_{eff}\right) 
		\end{eqnarray}
	The value of the deceleration parameter signifies whether a cosmic expansion is accelerated or not. For accelerating expansion, the necessary condition is $q<0$. Some recent experiments found that the present value of the deceleration parameter is $q\approx-0.810^{+0.1} _{-0.1} $, which indicates that our universe is going through an accelerated expansion phase.
 \section{Dynamical system analysis}\label{Sec3}
In this section, we are presenting a dynamical system analysis for $f(Q)$ gravity in the presence of a DBI scalar field. Dynamical system analysis helps in constructing and understanding cosmological models. It allows for the exploration of how different solutions evolve over time and can provide insights into the early universe, dark energy, and dark matter. To formulate the dynamical system, it is important to choose the variables carefully such that the cosmological parameters and the dynamical system can be written in terms of the variables. By following the available literature \cite{Pal:2019qch}\cite{doi:10.1142/S0217732315500091}\cite{doi:10.1142/S021827180600942X}\cite{PhysRevD.103.103521}, we have considered the following dynamical variable to construct the dynamical system 
 \\
	\begin{equation}\label{eq32}
		x=\frac{\Psi -2Q \Psi_{Q}}{6H^2}, \hspace{0.1cm}z=\frac{\nu \Dot{\phi}}{\sqrt{3\left( 1+\nu\right) }H},\hspace{0.1cm}u=\frac{\sqrt{V\left( \phi\right) }}{\sqrt{3} H}
	\end{equation}
	It is clear from the transformation \eqref{eq32} that the variable $x$ is associated with $F(Q)$ gravity, and $z,u$ are related to the scalar field. First we will write down the density  parameters like $\Omega_Q,\Omega_\phi$ in terms of the dimensionless variable as:\\
	\begin{equation}\label{eq33}
			\Omega_Q=x \hspace{0.2cm}\textrm{and}\hspace{0.2cm} \Omega_\phi=z^2+u^2.
	\end{equation}
	The total dark energy density is defined as
	\begin{eqnarray}\label{eq34}
	\Omega_d=\Omega_Q+\Omega_\phi=x+z^2+u^2
	\end{eqnarray} 
	Now the first field equation \eqref{eq22} is transformed as\\	
	\begin{eqnarray}\label{eq35}
		\Omega_M =1-\left( x+z^2+u^2\right) 
	\end{eqnarray}
	From the second field equation \eqref{eq23} we get \\
	\begin{eqnarray}\label{eq36}
		\frac{\Dot{H}}{H^2}=\frac{3}{2}\left\lbrace  \frac{ x-1-\frac{z^2}{\nu}+u^2 }{ 1+2 Q \Psi_{Q Q} +\Psi_{Q} }\right\rbrace  
	\end{eqnarray}
	In order to close the dynamical system we need a particular form of $\Psi\left( Q\right) $. In the literature\cite{doi:10.1142/S0218271823500621,PhysRevD.102.024057,NARAWADE2022101020,Vishwakarma_2023,Vishwakarma:2024qvw}, several forms of $\Psi\left(Q\right)$ are studied by researchers, such as exponential form, power-law form, log square-root form, logarithmic form, etc..In this paper, we have considered the power law and the exponential form of  $\Psi\left( Q\right)$ respectively, due to their mathematical simplicity, and one can always recover the Einstein Gr as a solution of these models by assigning some particular value to the constant parameter. 
	\subsection{Power-law model:$\Psi\left( Q\right) =  n Q^m$} 
	We start our analysis with the power law form $\Psi\left( Q\right) = n Q^m $. Here, $m,n$ are constant parameters, and $Q$ is a non-metricity scalar. From the form of $\Psi\left( Q\right) $, it is clear that for $m=1$ we get $F\left( Q\right) = \left( 1+n\right)  Q$, which is symmetric teleparallel equivalent to Einstein GR where $\left( 1+n\right)$ is acting as a rescaled gravitational constant. Therefore, throughout the dynamical analysis, we consider $m\neq1$.
	Now Eq\eqref{eq35} can be written in terms of the variable \eqref{eq32} as:
	\begin{eqnarray}\label{eq37}
	\frac{\Dot{H}}{H^2}=\frac{3}{2}\left\lbrace  \frac{ x-1-\frac{z^2}{\nu}+u^2 }{1-m x }\right\rbrace 
	\end{eqnarray}
	Using the expression in \eqref{eq36} we can write down  $\omega_{eff} \hspace{0.2cm} \textrm{and} \hspace{0.2cm} q$ as:
	\begin{eqnarray}
	\omega_{eff}=-1-\left\lbrace  \frac{ x-1-\frac{z^2}{\nu}+u^2 }{1-m x }\right\rbrace\label{eq38} \\
	q=-1-\frac{3}{2}\left\lbrace  \frac{ x-1-\frac{z^2}{\nu}+u^2 }{1-m x }\right\rbrace \label{eq39 a}
	\end{eqnarray}	
	Differentiating the variable \eqref{eq32} w.r.to N and using the conservation equation \eqref{eq17},\eqref{eq26},\eqref{eq27} and Klein-Gordon equation \eqref{eq18} and \eqref{eq37},we get the dynamical system as;\\
	\begin{eqnarray}\label{eq39}
		\frac{dx}{dN}=\left( m-1\right) x\left\lbrace  \frac{x-1-\frac{z^2}{\nu} +u^2}{2\left( 1- mx\right) }\right\rbrace \hspace{1cm}
			\end{eqnarray}
			\begin{multline}\label{eq40}
					\frac{dz}{dN}=-\frac{3}{2} z \left\lbrace \frac{ x-1-\frac{z^2}{\nu}+u^2 }{\left( 1-m x \right) }\right\rbrace - \frac{\sqrt{3\left( 1+\nu\right) } \lambda u^2}{2\nu}\\-3 z\left( \frac{1+\nu}{2\nu}\right) \hspace{3.9cm}
			\end{multline}
			\begin{eqnarray}\label{eq41}
			\frac{du}{dN}=\frac{\lambda \sqrt{3\left( 1+\nu\right) } u z}{2\nu} - 3u\left\lbrace \frac{ x-1-\frac{z^2}{\nu} +u^2}{2\left( 1-m x\right) }\right\rbrace 
			\end{eqnarray}
             \begin{multline}\label{eq42}
	\frac{d\nu}{dN}=\left( \nu-1\right)   \frac{\sqrt{3\left( \nu+1\right) }}{\nu} \left( \mu z -\frac{\lambda u^2}{z}\right) \\-\frac{3\left( \nu^2-1\right) }{\nu} \hspace{4.2cm}
             \end{multline}

	Here $N=\log a\left( t\right) $ is logarithmic time scale and $\lambda=\frac{V_{,\phi}}{V}$ and $\mu=\frac{f_{,\phi}}{f}$ are constant.\\
	We will investigate the physical properties and dynamical stability of our cosmological model through the critical / equilibrium points of the nonlinear dynamical system \eqref{eq39} - \eqref{eq42}. Each critical point represents a particular epoch in the cosmic timeline. To obtain the critical points, we set the right-hand sides of \eqref{eq39} -\eqref{eq42} to zero and solve the systems of equations for the dynamical variables $x,z,u,\nu$. The physical properties can be studied by calculating the values of energy densities,Eos parameters, and deceleration parameters for each critical point. The dynamical stability properties are analyzed through the signature of eigenvalues of the Jacobian matrix corresponding to each critical point, known as Linear stability theory (Hartman-Grobman Theorem). In the present analysis, the conventional existence criteria were taken into consideration, which implies that the critical points exist in real physical space and the matter density parameter is non-negative, satisfying the condition $0\leq\Omega_M\leq1$.In addition, the expression of $\nu$ imposed another constraint $\nu>0$.Hence the considerable phase space is expressed as : $\mathbb{B}=\left\lbrace \left( x,z,u,\nu\right) \in \mathbb{R}^4 :0\leq 1-x-z^2-u^2\leq1 \hspace{0.2cm}\textrm{and}\hspace{0.2cm} \nu>0\right\rbrace$. Following the above constraints, we have found a total of eight critical points listed in Table 1. A detailed analysis of each critical points are presented in the next paragraph.
\end{multicols}
\begin{table}[h!]
	\centering
	\begin{tabular}{|c|c|c|c|c|c|}
		\hline
	
		Critical point &x&z&u&$\nu$&Existence condition \\
	
		\hline
		\hline
		&&&&&\\
		$A_{1\pm}$&$\hspace{0.2cm}0\hspace{0.2cm}$&$-\frac{\sqrt{3}}{\sqrt{2} \lambda}$&$\pm\frac{\sqrt{3}}{\sqrt{2} \lambda}$&1&$\lambda\neq0$ \\
		\hline
		&&&&&\\
	    $A_{2\pm}$&$0$&$-\frac{\lambda}{\sqrt{6}}$&$\pm \frac{\sqrt{18+3\lambda^2-\lambda^4}}{\sqrt{6} \sqrt{3+\lambda^2}}$&1&$-\sqrt{6}<\lambda<\sqrt{6}$\\
		\hline
		&&&&&\\
	   $A_{3\pm}$&$0$&$\pm1$&$0$&$\frac{1}{3}\left( \mu^2-3\right) $ &$\mu^2 >3$\\
	
		\hline
&&&&&\\
	   $A_{4\pm}$&$0$&$\pm1$&$0$&1&Always\\
		\hline
	\end{tabular}
	\caption{Critical points along with their existence condition for model-$1$}
	\label{Table1}
\end{table}
\begin{multicols}{2}
	\begin{itemize}
	\item \textbf{Critical point $A_{1\pm}$ :} The value of energy densities  for these critical points are $\Omega_M=1-\frac{3}{\lambda^2}$ and $\Omega_d = \frac{3}{\lambda^2}$.Now, by applying the following constraint $0\leq\Omega_M\leq1$, we find that these two critical points will represent a valid cosmological solution for $\lvert \lambda\rvert \geq\sqrt{3}$.For $\lambda=\sqrt{3}$ , we have $\Omega_M=0$ and $\Omega_d=1$,which exhibits a complete dark energy domination. For $\lvert\lambda\rvert>\sqrt{6}$, we get $\Omega_M >\Omega_d$ represents that the dark matter is dominating over the dark energy component. Moreover, these two critical points represent a cosmological era in which the deceleration parameter $q$ has a constant value $\frac{1}{2}$  and effective Eos parameter  $\omega_{eff}=0$ implying a decelerated matter-dominated era. The eigenvalues of the Jacobian matrix for these critical points are
	$
	\begin{aligned}
	   &\left\lbrace \frac{1-m}{2},\frac{-3\left( \lambda+\mu\right) }{\lambda},\frac{3\left( -\lambda^3 +\sqrt{24\lambda^4-7\lambda^6}\right) }{4\lambda^3} , \right.\\
   & \left. \frac{3\left( -\lambda^3 +\sqrt{24\lambda^4-7\lambda^6}\right) }{4\lambda^3}\right\rbrace   
	\end{aligned}
	$   
 \begin{figure}[H]
		\centering
		\includegraphics[width=0.79\linewidth]{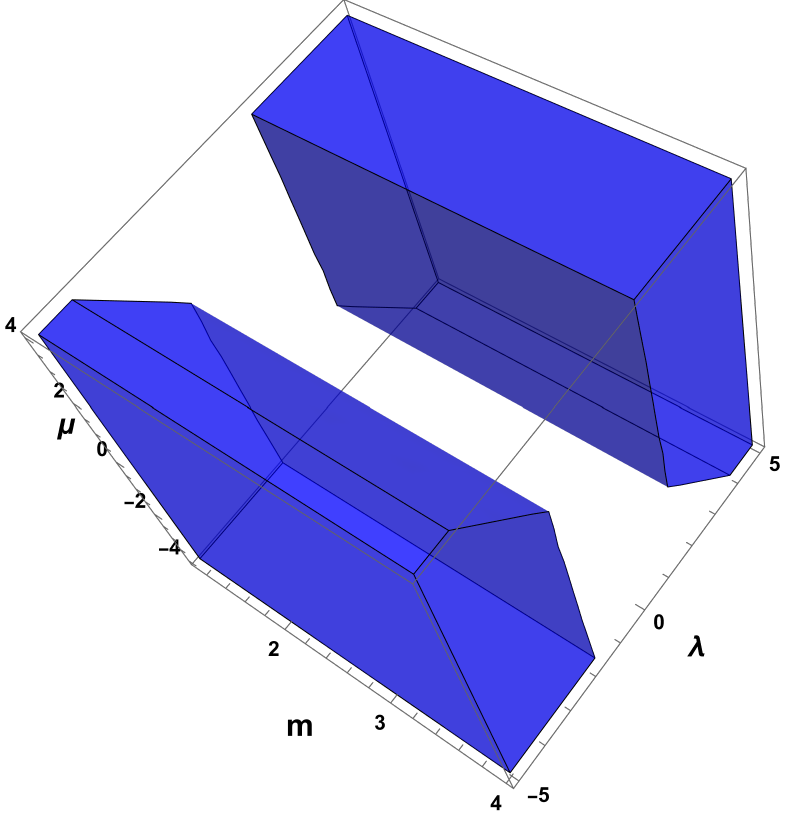}
		\caption{3 Dimensional region where $A_{1\pm}$ exhibits stable behavior}
		\label{fig1}
	\end{figure}
 \newpage
 Clearly, the eigenvalues and, hence, the stability depend on the values of $m,\lambda,\mu$. In Figure \ref{fig1}, we have presented a region in $ m-\lambda$ space, where these two critical points exhibit a stable behavior. Also, from the phase portrait \ref{fig4b}, it can be visualized that all the trajectories in the neighborhood of $A_{1+} \hspace{0.2cm}\textrm{and}\hspace{0.2cm} A_{1-}$ are converging toward these points, indicates the dynamical stability 
 \end{itemize}
	\begin{itemize}
	\item \textbf{Critical point $A_{2\pm}$:} At the critical points $A_{2+} \hspace{0.2cm}\textrm{and}\hspace{0.2cm}A_{2-}$, the value of energy densities are calculated as $\Omega_M=0\hspace{0.2cm}\textrm{and}\hspace{0.2cm}\Omega_d=1$.Therefore,  these two critical points represent a cosmological epoch completely dominated by dark energy. The value of the effective Eos parameter relies on the parameter $\lambda$ as $\omega_{eff}=\frac{\lambda^2-3}{3}$.Hence, it is clear that for several choices of parameter $\lambda$ can lead us to different evolutionary epochs. To identify the physically viable existence condition, we consider the expression under the square root is nonnegative, which imposes the constraint $-\sqrt{6}\leq\lambda\leq\sqrt{6}$ on the parameter. If we consider \\
   
    $\lvert\lambda\rvert<\sqrt{2}$,then the value of effective Eos parameter lying between $-1$ to $-\frac{1}{3}$ and the deceleration parameter $q<0$ , represents a accelerated quintessence scenario. If the value of scalar field potential is sufficiently small i.e $\lambda \to 0$, we get $\omega_{eff}\to-1$, represents an accelerated De-sitter phase where the dark energy component mimics the cosmological constant behavior. These two critical points can also represents decelerated matter era $\left( i.e\hspace{0.2cm} \omega_{eff}=0 \hspace{0.2cm} \textrm{and}\hspace{0.2cm} q=\frac{1}{2}\right)$  for $\lvert\lambda\rvert=\sqrt{3}$ and a stiff fluid solution for $\sqrt{6}\leq\lvert\lambda\rvert\leq\sqrt{3}$.To determine the dynamical stability, we compute the eigenvalues of the Jacobian matrix corresponding to these critical points as\\
	$\left\lbrace\frac{\lambda^2\left( 1-m\right) }{6},\frac{\lambda^2-6}{2},\lambda^2-3,-\lambda\left( \lambda+\mu\right) 
	\right\rbrace $
	\\
	In Figure \ref{fig2}, we have presented a possible region where $A_{2\pm}$ exhibits stable dynamical behavior. One can also visualized the trajectory around $A_{2\pm}$ in the phase portrait \ref{fig4a}, representing the same stable behavior
 \end{itemize}. 
 \end{multicols}
		\begin{figure}[H]
		\centering
		\includegraphics[width=0.45\linewidth]{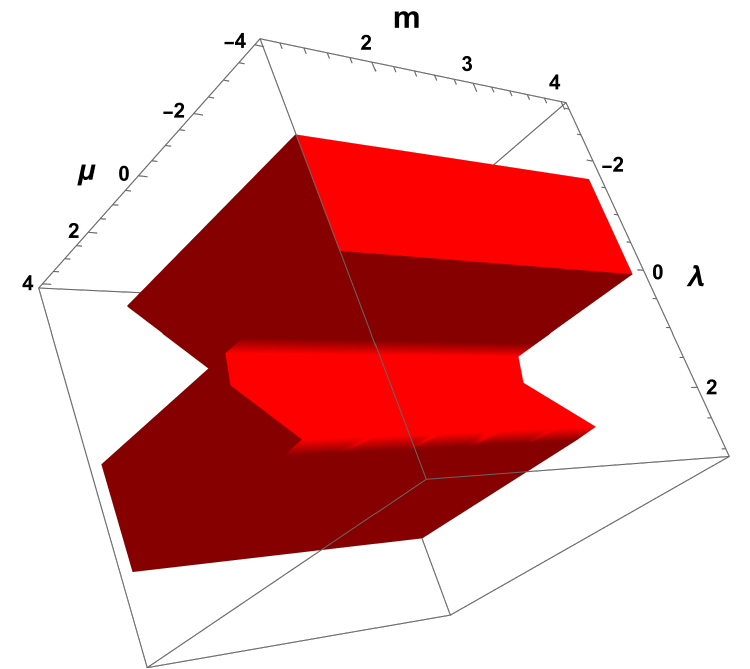}
		\caption{3 dimensional region where $A_{2\pm}$ exhibits stable bahavior.}
		\label{fig2}
	\end{figure}
	\begin{multicols}{2}
	\begin{itemize}
	\item 	\textbf{Critical point $A_{3\pm}$ :}The set of critical point $A_{3+} ,A_{3-}$ represents a valid cosmological solution for $\mu>\sqrt{3} \vee \mu<-\sqrt{3}$.The deceleration parameter and effective Eos parameter at these critical points are calculated as $q=\frac{6+\mu^2}{2\left( -3+\mu^2\right) }$ and $\omega_{eff}=\frac{3}{-3+\mu^2}$ respectively.For $\mu=\pm 2 \sqrt{3}$ , we get $\omega_{eff}=\frac{1}{3}$ and $q=1$ represents the radiation dominated decelerated phase of universe. Also, for $\mu=\sqrt{6}$, the deceleration parameter $q$ consist a value 2 and Eos parameter $\omega_{eff}=1$, indicates a decelerated stiff matter solution. Corresponding
    to these critical points, the Jacobian matrix has the set of eigenvalues \\
 $
 \begin{aligned}
 & \left\lbrace\frac{9}{-3+\mu^2},-\frac{\left( -1+m\right) \mu^2}{2\left( -3+\mu^2\right) },\frac{3\mu^2\pm 3\lambda\sqrt{\mu^2}}{2\left( -3+\mu^2\right) },\right.\\
 & \left.\frac{\pm\left(54\mu-9\mu^3+3\mu^5\right)-6\sqrt{\mu^2}(18+\mu^4)+36 \mu^3}{2\sqrt{\mu^2}\left(-3+\mu^2\right)^2} \right\rbrace    
 \end{aligned}	
 $
 \hspace{0.2cm}Under the constrain $\lvert\mu\rvert>\sqrt{3}$,the first eigenvalue $\frac{9}{-3+\mu^2}$ is always positive. Therefore, by linear stability theory, these two critical points cannot be stable. These two critical points can exhibit saddle or unstable behavior according to the value of the parameters $m,\lambda,\mu$.In Figure \ref{fig3}, a three-dimensional region is presented where $A_{3\pm}$ has
 saddle behavior. Also, in the phase diagram \ref{fig4a}, it is clear that some of the trajectories around $A_{3\pm}$ are attracted, and some of them are repelling, demonstrating a saddle character.	
\end{itemize}
\end{multicols}

 \begin{figure}[t]
		\centering
		\includegraphics[width=0.43\linewidth]{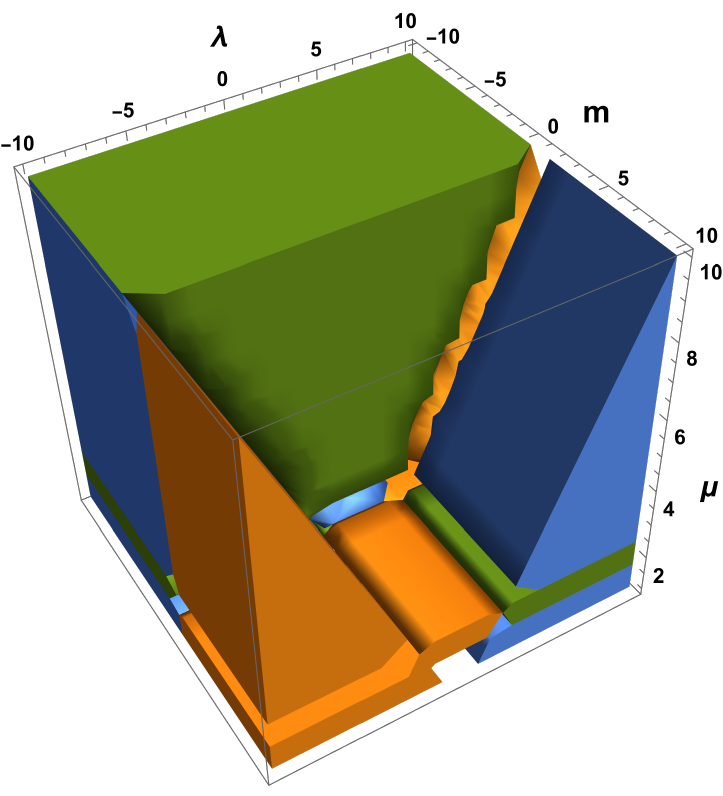}
		\caption{In \fcolorbox{black}{YellowOrange}{\rule{0pt}{2pt}\rule{2pt}{0pt}}\quad region only $A_{3+}$ exhibits saddle nature,in \fcolorbox{black}{MidnightBlue}{\rule{0pt}{2pt}\rule{2pt}{0pt}}\quad region only $A_{3-}$ exhibits saddle nature and in \fcolorbox{black}{LimeGreen}{\rule{0pt}{2pt}\rule{2pt}{0pt}}\quad region both $A_{3\pm}$ exhibit saddle nature.}
		\label{fig3}
	\end{figure}
\begin{multicols}{2}
\begin{itemize}
		\item 	\textbf{Critical point $A_{4\pm}$ :}The value of effective Eos parameter and deceleration parameter $\omega_{eff}\hspace{0.2cm}\textrm{and}\hspace{0.2cm}q$ are given by $\omega_{eff}=1\hspace{0.2cm}\textrm{and}\hspace{0.2cm}q=2$ respectively , represents a decelerated stiff matter universe.The eigenvalues of the Jacobian matrix corresponding to these critical points are\\
		$\left\lbrace 3,1-m,\frac{1}{2}\left(6+\sqrt{6}\lambda \right),-6+\sqrt{6}\mu\right\rbrace $
		\\
    Due to the presence of one positive eigenvalue, stability is not possible for these critical points. For $m>1\vee\lambda<-\frac{1}{\sqrt{6}}\vee\mu<\frac{1}{\sqrt{6}}$ , these critical points represents saddle character. The trajectories of phase space in the neighborhood of $A_{4\pm}$ corresponding to the saddle scenario are presented in Figure[\ref{fig4b}].\\
    \\
    The evolution of density parameters are presented in Figure[\ref{fig5}], where the vertical line corresponding to $N=0$ indicates the present cosmological epoch. As several cosmological projects revealed that the dark energy consists of $70\%$ and the remaining $30\%$ are dark matter. For Model-1 , we obtain the present values of $\Omega_M \approx0.3 \hspace{0.2cm}\textrm{and}\hspace{0.2cm} \Omega_d \approx 0.7$ respectively, which is compatible with observational data. Also, from the evolution of density parameter $\Omega_M$ and $\Omega_d$ presented in Figure \ref{fig5a},it is evident that in the early universe the matter density $\Omega_M$ dominates over the dark energy density $\Omega_d$. However, as the Universe evolves, the matter density $\Omega_M$ gradually decreases, while the dark energy density $\Omega_d$ becomes increasingly dominant. In the far future, the dark energy density completely surpass the matter density i.e $\Omega_M=\frac{\rho_M}{3H^2}\to 0$ and $\Omega_d\to 1$, leading to a Universe completely dominated by the dark energy sector. From Figure [\ref{fig5b}], we can observe that the evolution of the total Eos parameter ($\omega_{eff}$) begins with a matter era corresponding to $\omega_{eff}=0$ and then passes through the quintessence phase $\left( -1\le\omega\le0\right)$ and finally converges to de-sitter phase ($\omega_{eff}=-1$) in future i.e the late time evolution is similar to the $\Lambda$-CDM model. We can also observe that the present value of dark energy Eos parameter is obtained as $\omega_d \approx-1$, which is consistent with present Planck collaboration results $\left[ \omega_d\left( \textrm{at}\hspace{0.2cm} z=0\right) =-1.028\pm0.032\right] $. Furthermore, the evolution of the deceleration parameter presented in Figure[\ref{fig5b}], exhibits a transition from a decelerated $\left( q>0 \right) $ era to an accelerated $\left( q<0\right) $ era. The present value of the deceleration parameter is obtained as $q\approx-0.6$, which satisfies the observational evidence.
	\end{itemize}
 
\end{multicols}  
	\begin{figure}[H]
		\centering
		\begin{subfigure}{0.48\textwidth}
			\includegraphics[width=1\textwidth]{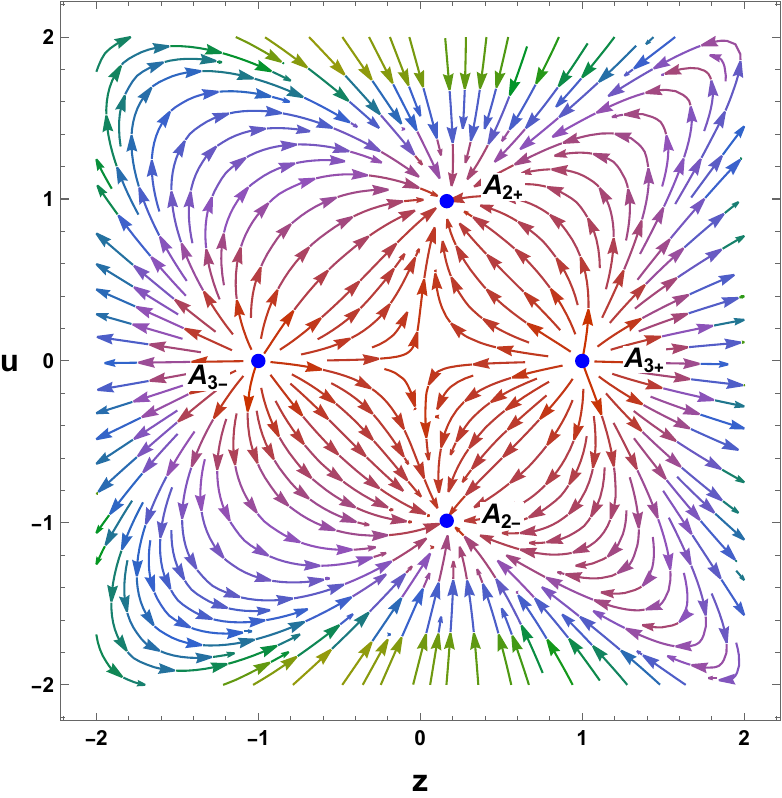}
			\caption{Projection of phase space diagram on $x_3-x_4$ plane around $A_{2\pm}$ and $A_{3\pm}$}
			\label{fig4a}
		\end{subfigure}
		\hfill
		\begin{subfigure}{0.49\textwidth}
			\includegraphics[width=1\textwidth]{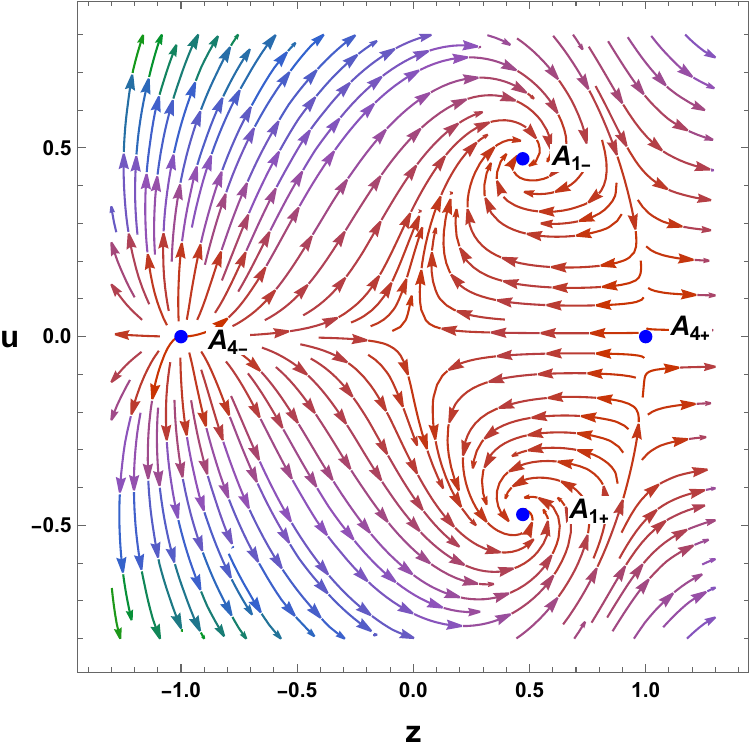}
			\caption{Projection of phase space diagram on $x_3-x_4$ plane around $A_{1\pm}$ and $A_{4\pm}$}
			\label{fig4b}
		\end{subfigure}
		
		\caption{Phase portrait of critical points for Model-1 }
		\label{fig4}
	\end{figure}
	\begin{figure}[H]
		\centering
		\begin{subfigure}{0.49\textwidth}
			\includegraphics[width=1\textwidth]{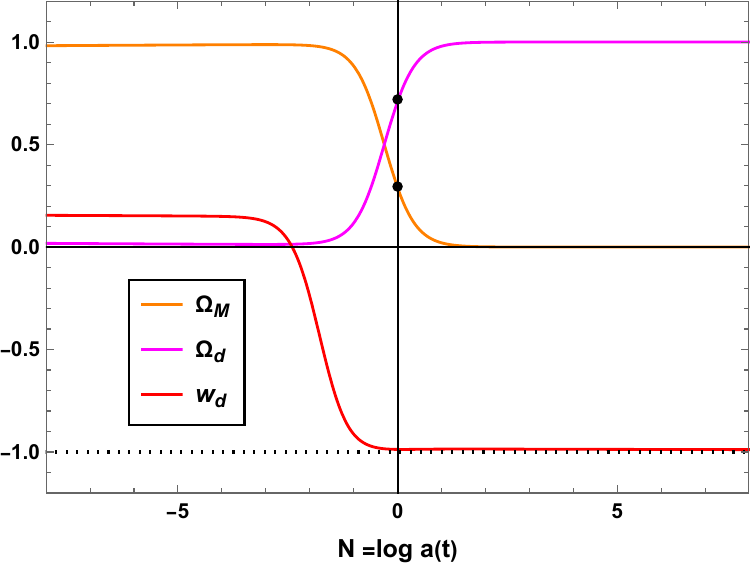}
			\caption{Evolution of density parameters $\Omega_M$,$\Omega_d$,$w_d$ for $m=1.16$,$\lambda=0.27$,$\mu=-0.98$ with fine-tuned initial condition}
			\label{fig5a}
		\end{subfigure}
		\hfill
		\begin{subfigure}{0.49\textwidth}
			\includegraphics[width=1\textwidth]{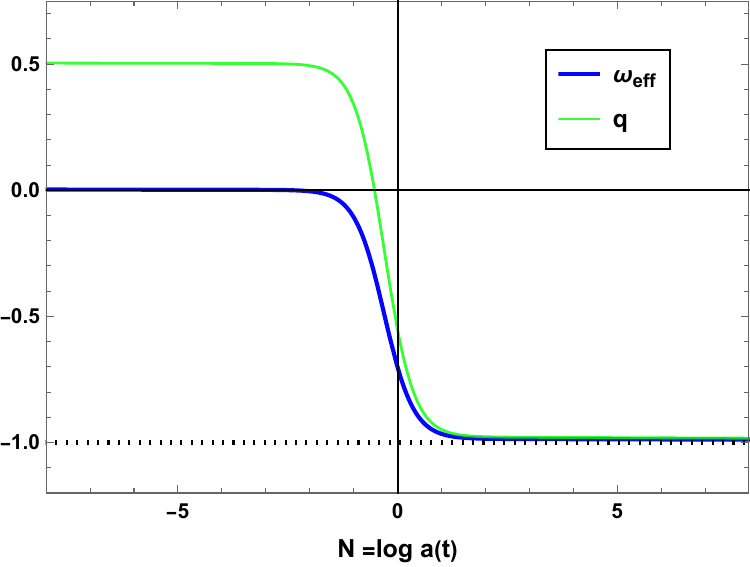}
			\caption{Evolution of  $\omega_{eff}$ and deceleration parameters $q$ for $m=1.16$ , $\lambda=0.27$ , $\mu=-0.98$ with fine-tuned initial condition}
			\label{fig5b}
		\end{subfigure}
		
		\caption{Evolution of cosmological parameters for Model-1}
		\label{fig5}
	\end{figure}

	\begin{multicols}{2}
 \subsection{Exponential Model :$\Psi\left( Q\right) =Q e^{\frac{\beta Q_0}{Q}}-Q$}
 In this subsection we consider the exponential form such as $\Psi\left( Q\right) =Q e^{\frac{\beta Q_0}{Q}}-Q$ , where $Q$ is usual nonmetricity term and $\beta$ is constant parameter and $Q_{0}=6 H_{0}^2$ with $H_0$ is current value of Hubble parameter $H$.For $\beta =0$, the cosmological model will be equivalent to Einstein GR without the cosmological constant. In order to construct the dynamical system, we have considered  the auxiliary variable as 
 \begin{multline}\label{eq44}
 	x=\frac{\Psi}{6H^2}\hspace{0.2cm},y=-2\Psi_{Q}\hspace{0.2cm},z=\frac{\nu \dot{\phi}}{\sqrt{3\left( 1+\nu\right) }H},\\
 	u=\frac{\sqrt{V\left( \phi\right) }}{\sqrt{3}H}\hspace{5cm}
 \end{multline}
 Under the above transformation, cosmological density parameters can be expressed as,
 \begin{eqnarray}\label{eq45}
 	\Omega_Q=x+y \hspace{0.2cm}\textrm{and}\hspace{0.2cm} \Omega_\phi=z^2+u^2
 \end{eqnarray}
 Therefore, the total dark energy density is given by
 \begin{eqnarray}\label{eq46}
 	\Omega_d=x+y+z^2+u^2
 \end{eqnarray}
 From the first field equation (\ref{eq22}), we can express the dark matter density in terms of the dynamical variable as
 \begin{eqnarray}\label{eq47}
 	\Omega_M=1-\left( x+y+z^2+u^2\right) 
 \end{eqnarray}
 Similarly, from the second field equation (\ref{eq23}) we get 
 \begin{eqnarray}\label{eq48}
 	\frac{\dot{H}}{H^2}=\frac{3\left( x+1\right)\left(x+y-1-\frac{z^2}{\nu} +u^2\right)  }{2\left( x+1\right)-y\left(x+1 \right)  +\left(2x+y \right)^2 }
 \end{eqnarray}
and the expressions for the total Eos parameter and deceleration parameter are 
\begin{eqnarray}
	\omega_{eff}=-1-\frac{2}{3}\left\lbrace \frac{3\left( x+1\right)\left(x+y-1-\frac{z^2}{\nu} +u^2\right)  }{2\left( x+1\right)-y\left(x+1 \right)  +\left(2x+y \right)^2 } \right\rbrace \\
	q=-1-\left\lbrace  \frac{3\left( x+1\right)\left(x+y-1-\frac{z^2}{\nu} +u^2\right)  }{2\left( x+1\right)-y\left(x+1 \right)  +\left(2x+y \right)^2 }\right\rbrace   \hspace{0.6cm}
\end{eqnarray}
Now differentiating the dynamical variable (\ref{eq44}) w.r.to $N$ and using the conservation equations (\ref{eq17}),(\ref{eq26}),(\ref{eq27}) and (\ref{eq48}) ,we get the dynamical system\\ 
\begin{eqnarray}
	\hspace{0.2cm}\frac{dx}{dN}=-\frac{3\left( x+1\right)\left(x+y-1-\frac{z^2}{\nu} +u^2\right)  }{2\left( x+1\right)-y\left(x+1 \right)  +\left(2x+y \right)^2 }\left( y+2x\right)\label{eq51}\\
	\frac{dy}{dN}=-\frac{3\left(x+y-1-\frac{z^2}{\nu}+u^2 \right) }{2\left( x+1\right)-y\left(x+1 \right)  +\left(2x+y \right)^2 }\left( y+2x\right) 
\end{eqnarray}
\begin{multline}
	\frac{dz}{dN}=-\frac{\sqrt{3\left(1+\nu \right) }}{2\nu}\left( \lambda u^2+\sqrt{3\left( 1+\nu\right) }z\right) \\
	-z \left( \frac{3\left( x+1\right)\left(x+y-1-\frac{z^2}{\nu} +u^2\right)  }{2\left( x+1\right)-y\left(x+1 \right)  +\left(2x+y \right)^2}\right)\hspace{0.2cm} 
\end{multline}
\begin{multline}
	\frac{du}{dN}=\hspace{0.4cm}\frac{\lambda\sqrt{3\left(1+\nu \right) }}{2\nu} u z\\
	-u\left(\frac{3\left( x+1\right)\left(x+y-1-\frac{z^2}{\nu} +u^2\right)  }{2\left( x+1\right)-y\left(x+1 \right)  +\left(2x+y \right)^2} \right)\hspace{0.2cm} 
\end{multline}
\begin{multline}\label{eq55}
	\frac{d\nu}{dN}=\left( \nu-1\right)\frac{\sqrt{3\left( \nu+1\right) }}{\nu} \left(\mu z-\frac{\lambda u^2}{{z}} \right) \\
	-\frac{3\left( \nu^2-1\right) }{\nu}\hspace{4.2cm}
\end{multline}
Here $N=\log a(t)$ and $\lambda=\frac{V_\phi}{V} \hspace{0.2cm} \textrm{and}\hspace{0.2cm}\mu=\frac{f_\phi}{f}$ are constant parameter. The critical points for the above system (\ref{eq51}) - (\ref{eq55}) are given in the table below.

\begin{figure}[H]
			\centering
			\includegraphics[width=0.5\linewidth]{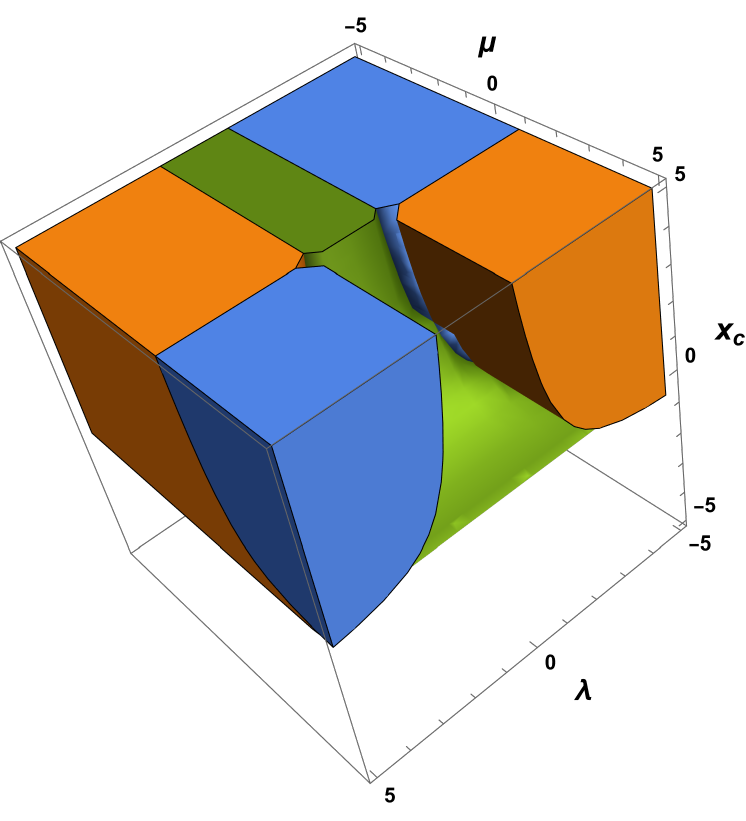}
			\caption{In \fcolorbox{black}{YellowOrange}{\rule{0pt}{2pt}\rule{2pt}{0pt}}\quad region only $B_{1+}$ exhibits saddle nature,in \fcolorbox{black}{NavyBlue}{\rule{0pt}{2pt}\rule{2pt}{0pt}}\quad region only $B_{1-}$ exhibits saddle nature and in \fcolorbox{black}{LimeGreen}{\rule{0pt}{2pt}\rule{2pt}{0pt}}\quad region both $B_{1\pm}$ exhibit saddle nature.}
			\label{fig6}
		\end{figure}

	\end{multicols}
	\begin{table}[h!]
		\centering
		\begin{tabular}{|c|c|c|c|c|c|c|}
			\hline
			
			Critical point &x&y&z&u&$\nu$&Existence  \\
			&&&&&&Condition\\
			\hline
			\hline
			&&&&&&\\
			$B_{1\pm}$&$\hspace{0.2cm}x_c\hspace{0.2cm}$&$-2 x_c$&$\pm \sqrt{1+x_c}$&$0$&1&$x_c\geq-1$ \\
			\hline
			&&&&&&\\
			$B_{2\pm}$&$x_c$&$-2x_c$&$-\frac{\sqrt{\frac{3}{2}}}{\lambda}$&$\pm \frac{\sqrt{\frac{3}{2}}}{\lambda}$&1&$\lambda\neq0$\\
			\hline
			&&&&&&\\
			$B_{3\pm}$&$\frac{3-\lambda^2 \pm\sqrt{9-6u_c^2\lambda^2}}{\lambda^2}$&$\frac{2\left(-3+\lambda^2-\left(\pm\sqrt{9-6u_c^2\lambda^2} \right)\right) }{\lambda^2}$&$-\frac{\sqrt{3}\pm\sqrt{3-2u_c^2\lambda^2}}{\sqrt{2}\lambda}$&$u_c$&$1 $ &$u_c^ 2\leq\frac{3}{2 \lambda^2}\wedge \lambda\neq0$\\
			&&&&&&\\
			
			\hline
			&&&&&&\\
			$B_{4\pm}$&$0$&$0$&$\pm 1$&$0$&$\frac{\mu^2 -3}{3}$&$\mu^2 \geq3$\\
			&&&&&&\\
			\hline
		\end{tabular}
		\caption{Critical points along with their existence condition for model-$2$}
		\label{Table1}
	\end{table}
	\begin{multicols}{2}
		\begin{itemize}
		\item 	\textbf{Critical point $B_{1\pm}$ :}The value of Eos parameter at the critical points $B_{1+} \hspace{.2cm} \textrm{and} \hspace{0.2cm}B_{1-}$ are $\omega_{eff}=1$.Hence, these are the two critical points representing a stiff matter-dominated solution. Also, the value of the deceleration parameter is positive $\left( q=2\right) $, indicating a decelerated expansion era. The Eigenvalues of the Jacobian matrix at these critical points are $
  \begin{aligned}
   & \left\lbrace0 \hspace{0.2cm},3\hspace{0.2cm},\frac{3\left(3+5x_c+2x_c^2 \right) }{\left( 1+x_c\right)^2 }\hspace{0.2cm},\right.\\
  & \left.\frac{6\pm\sqrt{6\left(1+x_c \right) }\lambda }{2}\hspace{0.2cm} , -\sqrt{6}\left( 1+\sqrt{1+x_c}\mu\right)  \right\rbrace   
  \end{aligned}	
	$
 \\
 Due to the presence of one vanishing eigenvalue, these two critical points are normally hyperbolic\cite{coley1999dynamicalsystemscosmology}. Also there is an eigenvalue with positive signature, implying that these two critical points exhibit unstable or saddle behavior depending on the values of parameter $\lambda,\mu$ and $x_c$.
		We have numerically evaluated a possible saddle region for these critical points in Figure[\ref{fig6}]
  
	\end{itemize}
	
\begin{figure}[H]
	\centering
	\begin{subfigure}{0.49\textwidth}
		\includegraphics[width=1\textwidth]{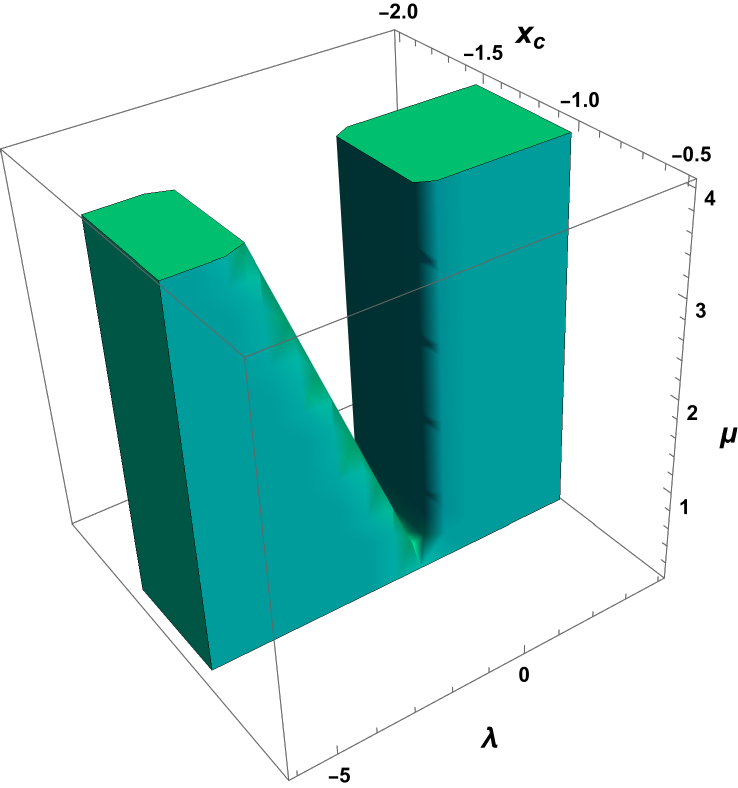}
		\caption{3 dimensional region where $B_{2\pm}$ exhibits  stable behavior}
		\label{fig7a}
	\end{subfigure}
	\hfill
	\begin{subfigure}{0.49\textwidth}
		\includegraphics[width=1\textwidth]{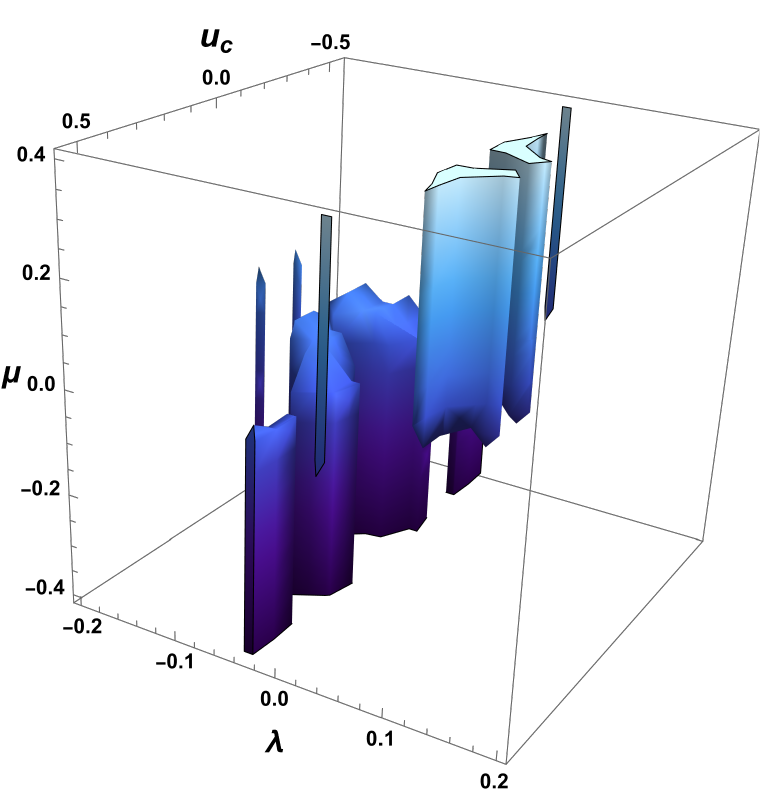}
		\caption{3 dimensional region where $B_{3-}$ exhibits  stable behavior}
		\label{fig7b}
	\end{subfigure}
	
	\caption{Region plots for critical points $B_{2\pm}$ and $B_{3-}$}
	\label{fig7}
\end{figure}

		\begin{itemize}
		\item 	\textbf{Critical point $B_{2\pm}$ :}These two critical points will exist in a real physical phase space until the potential of the Dbi field is nonconstant, i.e. $\lambda \neq 0$. The cosmological solution associated with the critical points $B_{2\pm}$ represents a decelerated matter-dominated phase of the universe since the values of effective Eos parameter and deceleration parameter are $\omega_{eff}=0$ and $q=1$ respectively. The eigenvalues of the Jacobian matrix corresponding to these two critical points are\vspace{0.2cm}
  
  $
  \begin{aligned}
    &  \left\lbrace 0\hspace{0.2cm},\frac{3\left(3+2x_c \right) }{2\left( 1+x_c\right) }\hspace{0.2cm},-\frac{3\left( \lambda+\mu\right) }{\lambda} \hspace{0.2cm},\right.\\
    &\left.-\frac{3}{4} -\frac{3\sqrt{\left( 1+x_c\right) \lambda^4\left( 24-7\lambda^2\left( 1+x_c\right)\right) }}{4\lambda^3\left( 1+x_c\right) }\right\rbrace 
  \end{aligned}		
  $
  \\

		Due to the existence of one vanishing eigenvalue, these two critical points are normally hyperbolic \cite{coley1999dynamicalsystemscosmology}. Therefore, their stability nature can be determined by the signature of non-vanishing eigenvalues. We have numerically evaluated a region in Figure[\ref{fig7a}] for some value combination of $\lambda$,$\mu$, and $x_c$ such that these two critical points may exhibit stable behavior. We have also presented the phase space trajectories around $B_{2\pm}$ in Figure[\ref{fig10b}], corresponding to the stable scenario.\\		
	\end{itemize}
 \end{multicols}
 
 \begin{multicols}{2}

		\begin{itemize}
		\item 	\textbf{Critical point $B_{3\pm}$ :}At these two critical points the dark energy density has a constant value $\Omega_d=1$ and the matter density $\Omega_M=0$, representing a completely dark energy dominated cosmological epoch. These two critical points will represent a valid cosmological scenario if the expression under the square root is non-negative. The value of effective Eos parameter and deceleration parameter is calculated as $\omega_{eff}=\frac{\pm\left(3-2u_c^2 \mu^2 \right)+\sqrt{9-6u_c^2\lambda^2} }{\pm3+\sqrt{9-6u_c^2 \lambda^2}}$ and $q=\frac{\pm\left( 6-3u_c^2 \lambda^2\right) +2\sqrt{9-6u_c^2\lambda^2}}{\pm3+\sqrt{9-6u_c^2\lambda^2}}$. Since the Eos and deceleration parameters are influenced by $\lambda$ and auxiliary variable $u$, several combinations of $\left(\lambda,u \right) $ can manifest different cosmological epochs commencing from the matter era to the late-time de-sitter era. By analyzing numerically, we have found that both the $\omega_{eff}\hspace{0.2cm}\textrm{and}\hspace{0.2cm}q$ corresponding to $B_{3+}$  consists only positive value for all possible combination of $\left(\lambda,u \right) $, which implies $B_{3+}$ characterized a decelerated solution and not represent the current cosmological scenario but capable to represents the radiation and matter era. On the other hand, the cosmological scenario associated with $B_{3-}$ is interesting since the effective Eos parameter and deceleration parameter can assume values in the interval $\left(-1,0 \right) $ according to the choices of parameter $\lambda$ and auxiliary variable $u$ respectively.The range of values of $\omega_{eff}$ that can be obtained from critical points  $B_{3-}$ is given in Figure[\ref{fig8}]. Hence, $B_{3-}$ can represent matter, quintessence, and de-sitter epoch, respectively, for proper choices of $\lambda$ and $u$.
  \begin{figure}[H]
			\centering
			\includegraphics[width=1.25\linewidth]{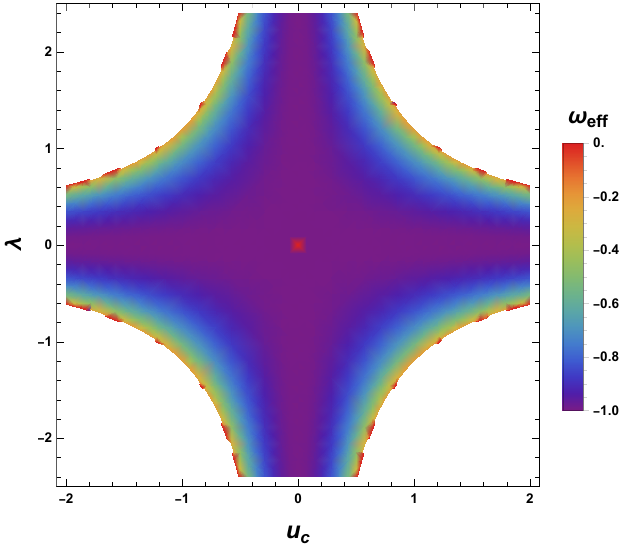}
			\caption{Variation of $\omega_{eff}$ for critical points $B_{3-}$}
			\label{fig8}
		\end{figure}

		\end{itemize}
  
  \end{multicols}
  \begin{figure}[H]
		\centering
		\begin{subfigure}{0.49\textwidth}
			\includegraphics[width=1\textwidth]{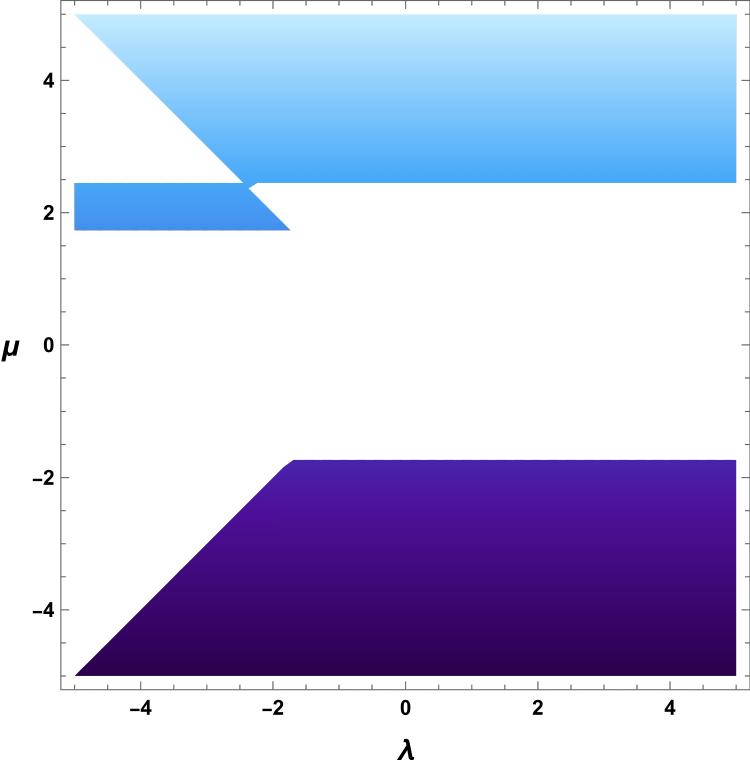}
			\caption{Possible region where $B_{4+}$ has saddle behaviour }
			\label{fig9a}
		\end{subfigure}
		\hfill
		\begin{subfigure}{0.49\textwidth}
			\includegraphics[width=1\textwidth]{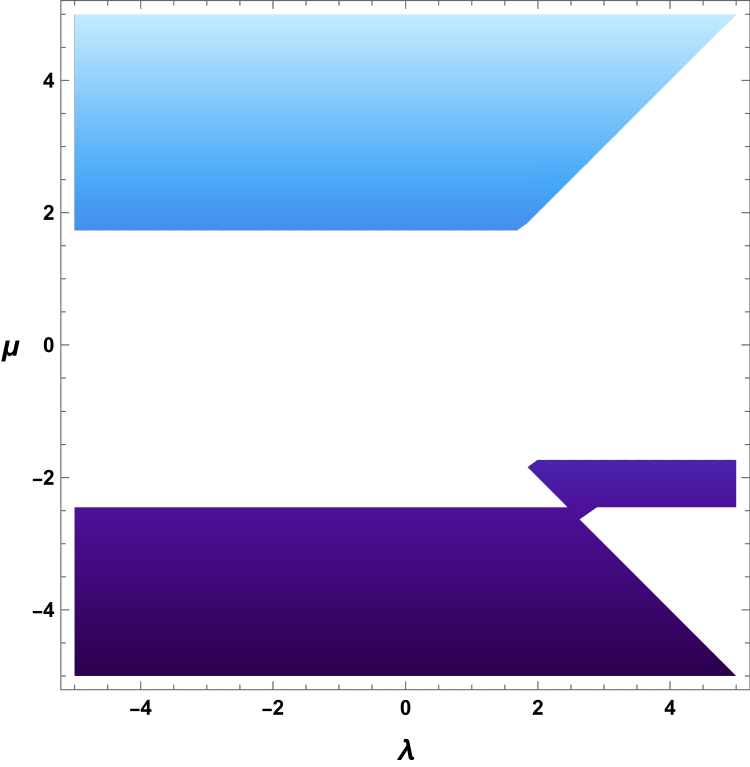}
			\caption{Possible region where $B_{4-}$ has saddle behaviour}
			\label{fig9b}
		\end{subfigure}
		
		\caption{Region plots for critical points $B_{4\pm}$ }
		\label{fig9}
	\end{figure}
 \begin{multicols}{2}
    The eigenvalues of the Jacobian matrix are too complex to be written in the manuscript. By investigating the eigenvalues numerically, we have found that $B_{3+}$ cannot be stable for any combination of $\lambda$, $\mu$, and $u_c$, but stability is possible for $B_{3-}$. The numerically evaluated stable region corresponding to $B_{3-}$ is presented in Figure[\ref{fig7b}].\\
    \begin{itemize}
        \item \textbf{Critical points $B_{4\pm}$:}The fourth set of critical points $B_{4\pm}$ characterized a cosmological solution that is influenced by the DBI-essence scalar field. These two critical points always exist in the real physical phase space; the only constraints, i.e., $\mu\geq\sqrt{3}\vee\mu\leq-\sqrt{3}$  arise from the non-negativity of the Lorentz boost factor. The value of density parameters corresponding to these critical points are $\Omega_M=0$ and $\Omega_d=1$, respectively, representing a dominant nation of dark energy components over matter. The value of the effective Eos parameter is obtained as $\omega_{eff}=\frac{3}{\mu^2-3}$.Under the above-mentioned physical constraints $\left(\mu^\geq3\right)$, these critical points cannot exhibit the accelerated solution, but for $\mu=\pm2\sqrt{3}$, we get $\omega_{eff}=\frac{1}{3}$ which represents the radiation era. The eigenvalues of the Jacobian matrix associated with these critical points are given by \\ 
  $
  \begin{aligned}
    &  \left\lbrace  0,\frac{9}{-3+\mu^2}\hspace{0.2cm},\frac{3\mu^2\pm 3\lambda\sqrt{\mu^2}}{2\left(-3+\mu^2\right)},\frac{9\mu^2}{2\left(-3+\mu^2\right)},\right.\\
    &\left.\frac{\pm\left(54\mu-9\mu^3+3\mu^5\right)-6\sqrt{\mu^2}(18+\mu^4)+36 \mu^3}{2\sqrt{\mu^2}\left(-3+\mu^2\right)^2}\right\rbrace 
  \end{aligned}		
  $
  Due to the presence of one vanishing eigenvalue, these two critical points are normally hyperbolic. Hence, the nature of stability depends on other non-vanishing eigenvalues. Under the physical constraint, the second eigenvalue is always positive, and therefore, stability is not possible. In Figure[\ref{fig9a}] and [\ref{fig9b}], we have presented a numerically evaluated region where $b_{4+}$ and $B_{4-}$ can exhibit saddle characteristics, respectively.The trajectories in the neighborhood of critical points $B_{4+}$ and $B_{4-}$ are presented in the phase diagram Figure[\ref{fig10b}], which illustrates the saddle behavior.\\
  \\
  The evolution of energy densities for the exponential model is illustrated in Figure[\ref{fig11a}]. One can note that we have obtained the values for energy densities corresponding to the exponential $f(Q)$ model as $\left(\Omega_M,\Omega_d \right)\approx\left(0.3,0.7 \right)$, which is very much identical to the observational result. The evolution of energy densities of the exponential model is similar to the power law model. i.e., in late time, the matter density $\Omega_m$ will be completely dominated by the dark energy density $\Omega_d$ as $\Omega_M \to 0$ and $\Omega_d\to1$.From the evolution of dark energy Eos parameter $\omega_d$ in Figure[\ref{fig11a}], we can notice that the present value of  $\omega_d$ is -1.01, which is consistent with the current cosmological data and successfully manifests the phantom behavior.
    \end{itemize}
    
 \end{multicols}

	\begin{figure}[H]
		\centering
		\begin{subfigure}{0.48\textwidth}
			\includegraphics[width=1\textwidth]{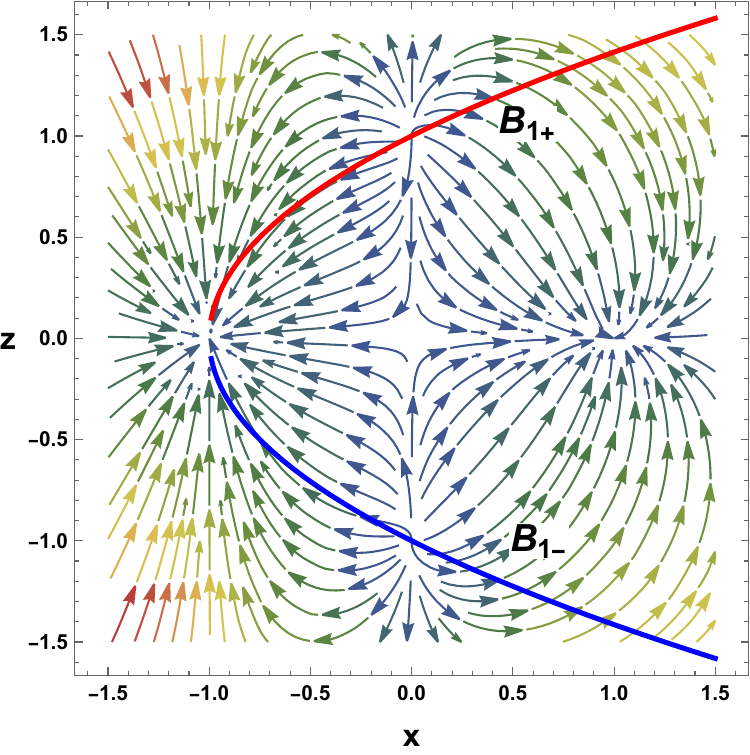}
			\caption{Projection of phase space on $x-z$ plane Red colored curve represents the family of critical points $B_{1+}$ and Blue colored curve represents the family of Critical points $B_{1-}$}
			\label{fig10a}
		\end{subfigure}
		\hfill
		\begin{subfigure}{0.485\textwidth}
			\includegraphics[width=1\textwidth]{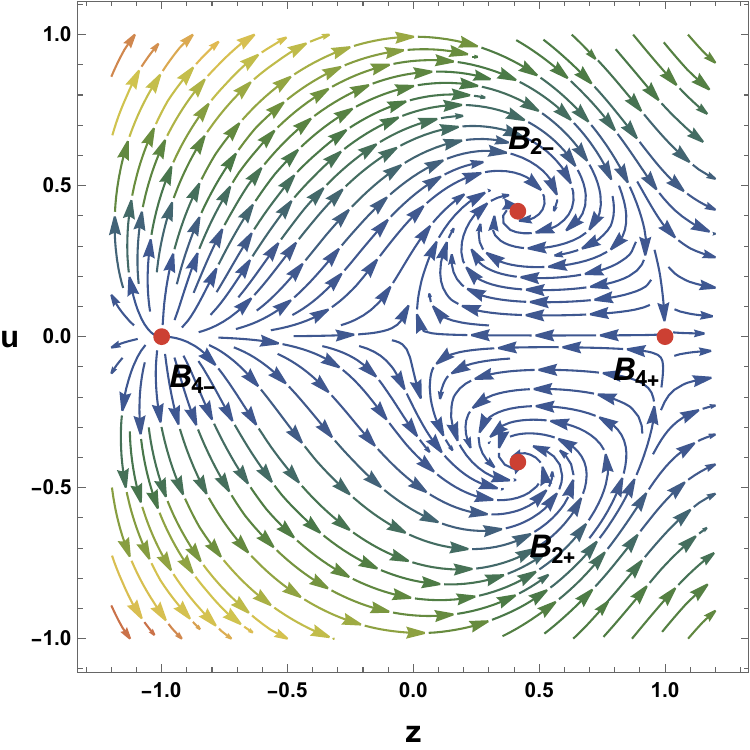}
			\caption{Projection of phase space on $z-u$ plane in the neighborhood of critical points $B_{2\pm}$ and $B_{4\pm}$\\}
			\label{fig10b}
		\end{subfigure}	
		\hfill
  \vspace{1cm}
		\begin{subfigure}{0.485\textwidth}
			\includegraphics[width=1\textwidth]{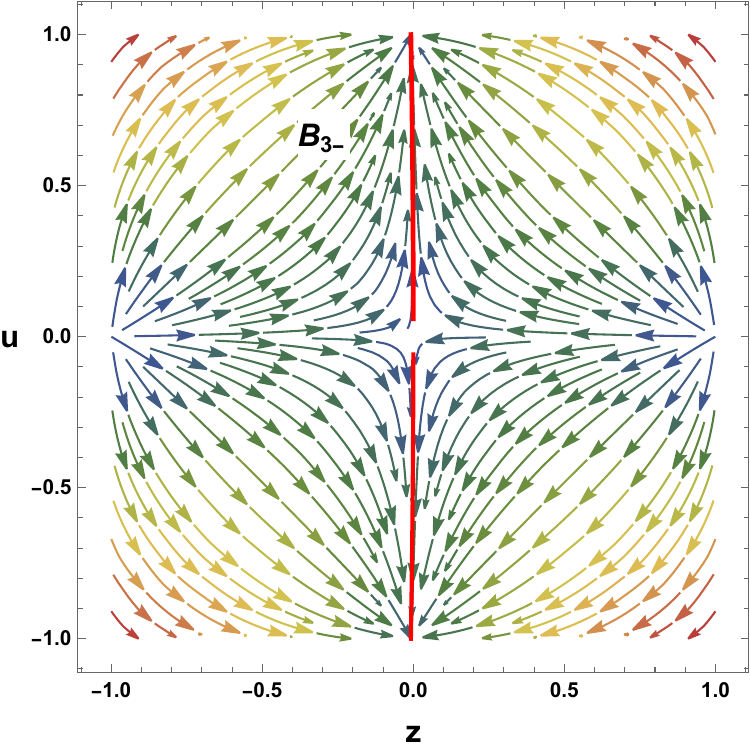}
			\caption{Projection of phase space on $z-u$ plane where, Red colored curve represents the family of critical points $B_{3-}$}
			\label{fig10c}
		\end{subfigure}
		\hfill
		\begin{subfigure}{0.485\textwidth}
			\includegraphics[width=1\textwidth]{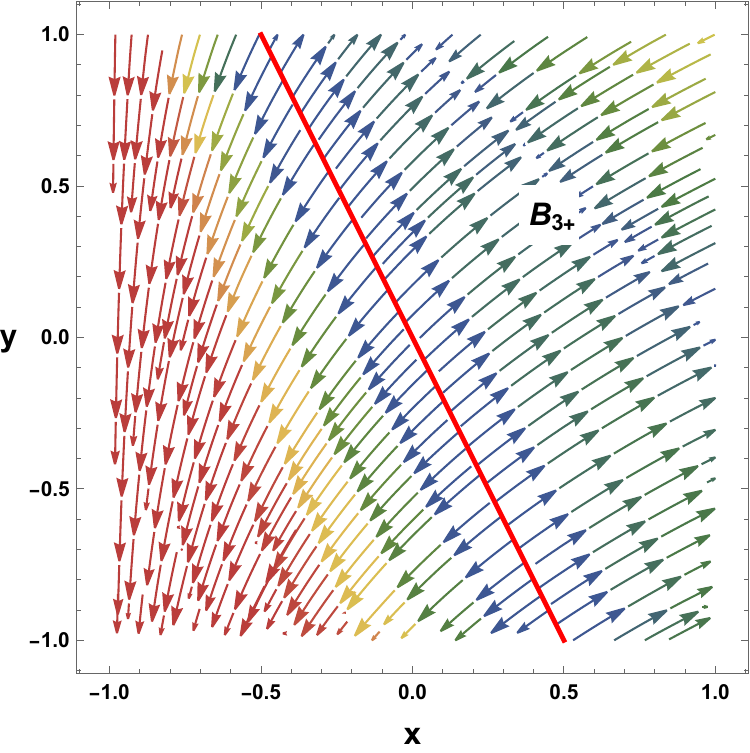}
			\caption{Projection of phase space on $x-y$ plane where, Red colored curve represents the family of critical points $B_{3+}$ }
			\label{fig10d}
		\end{subfigure}	
		\caption{Phase portrait of critical points for Model-2}
		\label{fig10}
	\end{figure}
 \begin{multicols}{2}
   On the other hand, the evolution curve of total Eos parameter $\omega_{eff}$ in Figure [\ref{fig11b}] depicts the main epochs of the cosmic timeline, starting from stiff matter era for $\omega_{eff}=1$, passing through decelerated radiation era $\omega_{eff}=\frac{1}{3}$ to matter era $\omega_{eff}=0$ respectively and then facing an accelerated quintessence era and finally converging to a de sitter phase in late time. Additionally, the evolution of the deceleration parameter $q$ is presented in Figure[\ref{fig11b}].As it can seen from Figure[\ref{fig11b}], the current value of the deceleration parameter corresponding to the exponential $f(Q)$ gravity model is obtained as $q_0\approx-0.5$, which satisfies the current observation results. The negative value of the deceleration parameter indicates the cosmological acceleration as a fundamental cosmological phenomenon.
 \end{multicols}

\begin{figure}[H]
	\centering
	\begin{subfigure}{0.47\textwidth}
		\includegraphics[width=1\textwidth]{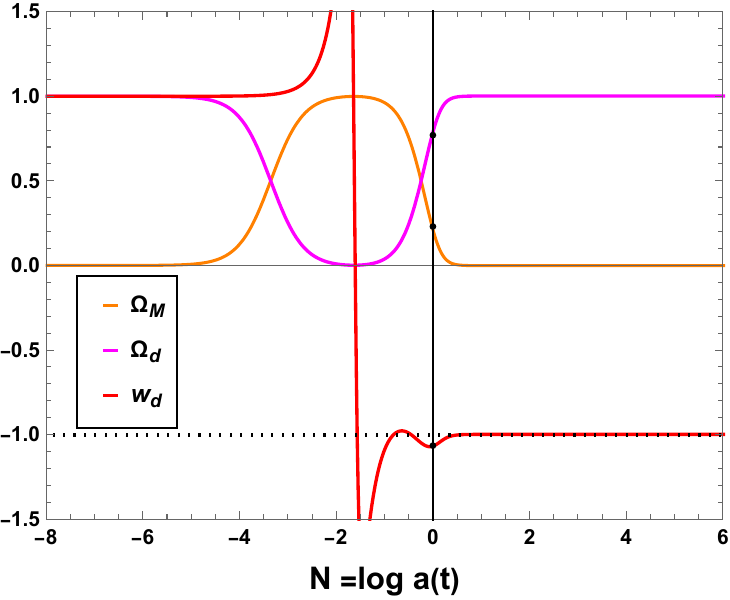}
		\caption{Evolution of density parameters $\Omega_M$,$\Omega_d$ and dark energy Eos parameter $\omega_d$ for $\lambda=0.27$, $\mu=-0.98$ with fine-tuned initial condition}
		\label{fig11a}
	\end{subfigure}
	\hfill
	\begin{subfigure}{0.47\textwidth}
		\includegraphics[width=1\textwidth]{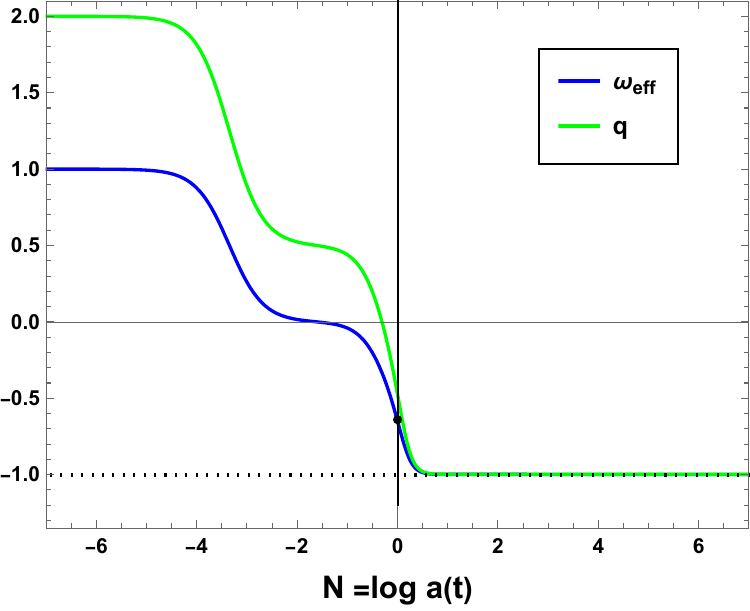}
		\caption{Evolution of effective equation of state $\omega_{eff}$ and deceleration parameters $q$ for   $\lambda=0.27$ , $\mu=-0.98$ with fine-tuned initial condition}
		\label{fig11b}
	\end{subfigure}
	
	\caption{Evolution of cosmological parameters for Model-2}
	\label{fig11}
\end{figure}
\begin{multicols}{2}

\section{Conclusions}\label{sec4}
 In the present paper, we have studied the dynamics of coincident $f(Q)$ gravity in the presence of a generalized DBI Essence scalar field through dynamical system analysis. The flat, homogenous FLRW metric describes the geometrical background of our cosmological setup. We have considered two different forms of $f(Q)$ gravity, namely, the power law model:$f(Q)=Q+nQ^m$ and the exponential form: $f(Q)=Q e^{\frac{\beta Q_0}{Q}}$ respectively, here $Q$ is the nonmetricity and $(\beta,n,m)$ are constant parameter. The DBI essence acts as an additional dark energy component with an exponential form of potential and warp factor. After presenting the field equations for the above-mentioned cosmological model, we employed the dynamical variable to construct the nonlinear dynamical system. An ideal cosmological model should represent some of the following cosmological eras: Inflation $\to$ Radiation era$\to$
 Matter era$\to$ late time acceleration era. For the power law model, we obtain a total of eight numbers of critical points, namely $A_{i\pm},i=1(1)4$. These sets of critical points can exhibit several cosmological epochs, starting from the decelerated stiff matter era to the late time acceleration era, according to the value combination of cosmological parameters.\\
 The critical points $A_{1\pm}$ represent a decelerated matter era with constant value of state parameter and deceleration parameter $\left(\omega_{eff},q\right)=\left(0,\frac{1}{2}\right)$. The dynamical stability of these two critical points depends on $m,\lambda,\mu$. The critical points $B_{2\pm}$ can recover several cosmological epochs according to the parameter $\lambda$, which is associated with the scalar field potential. For $\lvert\lambda\rvert<\sqrt{2}$, it can depict an accelerated quintessence era, and for $\lvert\lambda\rvert=\pm\sqrt{3}$ decelerated matter-dominated epoch can be recovered. Moreover, for small potential $\left(\lambda\to 0\right)$, the accelerated de-sitter era can be retrieved. The dynamical stability features are highly sensitive to cosmological parameters. For the critical points $A_{3\pm}$ ,the effective eos parameter $\omega_{eff}$ and deceleration parameter $q$ depends on parameter $\mu$ . Under the constraint of a positive warp factor, these two critical points do not represent the current accelerating era, while a radiation-dominated epoch can be recovered for $\mu=\pm 2\sqrt{3}$. Also, dynamical stability is not possible for these critical points under the constraint mentioned above. Finally, the set of critical points $A_{4\pm}$ is associated with a stiff matter solution, which is saddle or unstable in nature. \\
 In the exponential $f(Q)$ gravity model, we have a total of eight critical points, namely, $B_{i\pm},i=1(1)4$. The cosmological solution associated with $B_{1\pm}$ describes a decelerated stiff matter solution, which is dynamically saddle or unstable. The critical points $B_{2\pm}$ are associated with a decelerated matter-dominated solution. The third set of critical points $B_{3+}$ can represent a matter-dominated epoch or radiation era according to the parameter $\lambda$ and variable $u$. However, accelerated expansion cannot be achieved from this critical point. Moreover, the set of critical points $B_{3+}$ is interesting in terms of the current cosmological scenario. This solution is completely dark energy dominated, and the value of the effective Eos parameter and deceleration parameter lies in the range of $[-1,0 ]$. Hence, a late-time accelerated scenario can be identified for particular choices of $\lambda$ and $u$.Finally, the fourth set of critical points $B_{4\pm}$ exhibits a dark energy-dominated solution, which can be associated with the decelerated radiation era for a specific value of parameter $\mu$. \\
 From some current astronomical observations, researchers have found the value of the Eos parameter as :$\omega =-1.073^{+0.090}_{-0.089} \left(\textrm{WMAP+CMB}\right)$, $\omega=-1.035^{+0.055}_{-0.059} \left(\textrm{Supernovae Cosmogical Project}\right)$ and $\omega=-1.03^{+0.03}_{-0.03}\left(\textrm{Planck 2018}\right)$.In our present study, from the evolution of the Eos parameter in Figure [\ref{fig5}], the current value of the dark energy Eos parameter for the power-law model is obtained as $\omega_d\approx-1$, which is compatible with the observational result. Similarly, from Figure[\ref{fig11}], one can note that the value of the dark energy Eos parameter corresponding to the exponential model is $\omega_{d}\approx-1.01$. For both models, the late-time behavior of the Eos parameter is similar to the $\Lambda CDM$ model, and the exponential model exhibits crossing of the phantom divide line in the present cosmological epoch, which is more acceptable in terms of current observational results. Moreover,  the evolution of the deceleration parameter in both the power law and the exponential model reveals acceleration as a fundamental cosmological phenomenon that our universe is going through.\\
 From the phase space analysis of cosmological models presented in this paper, it is clear that the cosmological solutions associated with the critical points are useful in explaining the history of cosmological evolution. Additionally, the accelerated cosmological solution obtained for specific choices of the cosmological parameters can be considered as an important alternative to the $\Lambda CDM$ dark energy model. An extension of the present work can be done by considering different $f(Q)$ gravity models, such as the log square root model and the logarithmic model. Moreover, several types of interactions between dark sectors can provide a deeper understanding of cosmological evolution. \\
The analysis of dynamical systems in the context of coincident symmetric teleparallel $f(Q)$ gravity, especially when combined with a DBI-essence scalar field, is crucial for several reasons. By identifying critical points in the dynamical system, one can determine the stability and nature of these points (e.g., attractors, repellers, or saddle points). This helps in understanding the long-term behavior of the universe. Dynamical system analysis in symmetric teleparallel $f(Q)$ gravity with a DBI-essence scalar field provides a powerful tool for understanding the complex behaviors and potential physical implications of these theories. It aids in exploring the stability, evolution, and viability of cosmological models, thereby contributing to our understanding of the universe's dynamics and the fundamental nature of gravity.\\

\section*{\normalsize\bf{ACKNOWLEDGEMENTS}}
RM is thankful to UGC, Govt. of India,
for providing Senior Research Fellowship
(NTA Ref. No.: 211610083890). A. Pradhan is grateful for the facilities provided by IUCAA, Pune, India, through a visiting associateship.        
\section*{\normalsize\bf{Data Availability Statememt}}
We have developed our research paper completely in an analytical approach. We did not produce any data for this publication. As a result, our research is not associated with any one form of data.

 \printbibliography
\end{multicols}

\end{document}